\documentclass[aps,prl,floatfix,twocolumn,superscriptaddress,longbibliography]{revtex4-1}
\usepackage[english]{babel} 
\usepackage[latin1]{inputenc}
\usepackage[usenames,dvipsnames]{color}

\usepackage{amsmath,amssymb,amsfonts}
\usepackage[colorlinks=true,linkcolor=blue,urlcolor=blue,citecolor=blue]{hyperref}
\usepackage[percent]{overpic}
\usepackage[latin1]{inputenc}
\usepackage{amsmath,amssymb}
\usepackage{graphicx,color}
\usepackage{braket}
\usepackage{bm}
\usepackage[caption = false]{subfig}

\usepackage{titlesec}
\titleformat{\section}[runin]{\normalfont\itshape}{}{3pt}{}[.]

\usepackage[resetlabels,labeled]{multibib}
\newcites{Fig}{References of Figures}

\newcommand{\be}{\begin{eqnarray}}
\newcommand{\ee}{\end{eqnarray}}

\renewcommand{\vec}[1]{\bm{#1}}

\definecolor{green(html/cssgreen)}{rgb}{0.0, 0.5, 0.0}

\begin{document}
\title{Fractional spin excitations and conductance in the spiral staircase Heisenberg ladder}

\author{Flavio Ronetti}
\affiliation{Department of Physics, University of Basel, Klingelbergstrasse 82, CH-4056 Basel, Switzerland}
\author{Daniel Loss}
\affiliation{Department of Physics, University of Basel, Klingelbergstrasse 82, CH-4056 Basel, Switzerland}

\author{Jelena Klinovaja}
\affiliation{Department of Physics, University of Basel, Klingelbergstrasse 82, CH-4056 Basel, Switzerland}

\begin{abstract} 
We investigate theoretically the spiral staircase Heisenberg spin-$1/2$ ladder in the presence of antiferromagnetic long-range spin interactions and a uniform magnetic field. As a special case we also consider the Kondo necklace model.
If the magnetizations of the two chains forming the ladder satisfy a certain resonance condition, involving interchain couplings as perturbations, the system is in a partially gapped magnetic phase hosting excitations characterized by fractional spins, whose values can be changed by the magnetic field. 
We show that these fractional spin excitations can be probed via the magnetization and by spin currents in a  transport setup with a spin conductance that reveals the fractionalized spin. In some special cases, the spin conductance reaches universal values. We obtain our results analytically via bosonization techniques as well as numerically via density matrix renormalization group methods and find remarkable agreement between the two approaches.
\end{abstract}

\maketitle
\textit{Introduction.}  In conventional magnetic systems, typical excitations are collective spin waves that can be quantized in terms of massless bosons with integer spin $S=1$, called magnons~\cite{Morimae05,Demokritov06,Serga14,Bozhko16,Nakata17,EdNatMag,Cramer18}. These quanta of spin waves have been intensively investigated in relation to information transport~\cite{Kajiwara10,Cornelissen15,Wimmer20}, topology~\cite{Hoogdalem13,ZhangL13,Shindou13,Shindou13b,Mook14,Mook16rev,Nakata17topo,McClarty21,Mook21}, and control of spin textures~\cite{Yan11,Wang12,Psaroudaki18}. When quantum fluctuations are relevant, as is the case in one-dimensional spin-$1/2$ systems, even more fascinating excitations can arise~\cite{Haldane80,Giamarchi03,Lake05}. A paradigmatic example is the antiferromagnetic Heisenberg spin-$1/2$ chain~\cite{Heisenberg28,Haldane83,Meier01}, where the excitations are spinons, i.e. propagating domain walls with spin $1/2$~\cite{Haldane91,Karbach97,Lake00,Zheludev00,Bera17}. 
From the experimental point of view, it is believed that many compounds can be modeled as Heisenberg chains or ladders~\cite{Nagler91,Tennant95,Ruegg08,Ward17}, where spinons have been observed~\cite{Thielemann09,Lake10,Mourigal13}.

\begin{figure}[h]
	\centering
	\includegraphics[width=0.87\linewidth]{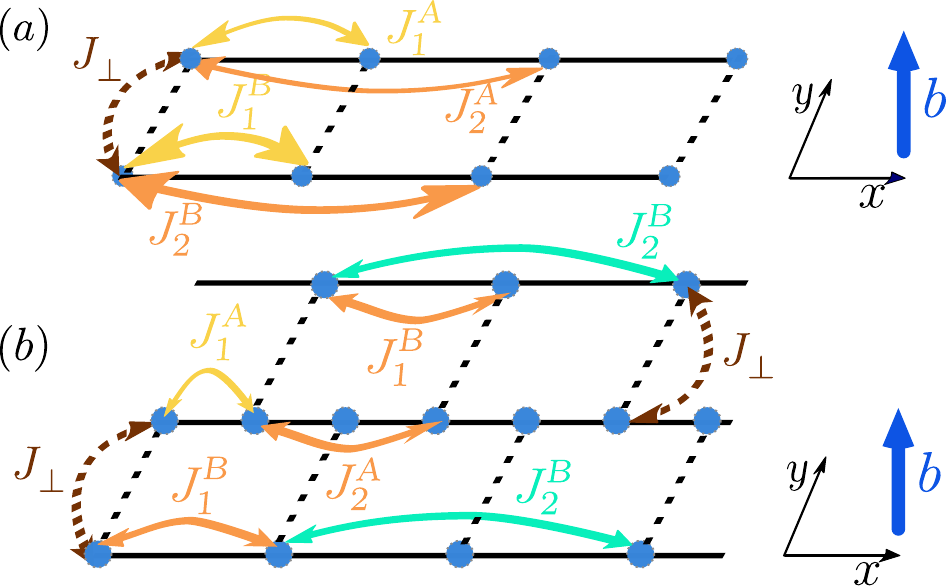}
	\caption{(a) 
		The spiral staircase Heisenberg ladder is composed of two weakly-coupled  spin-$1/2$ (blue dots) chains aligned along the $x$ direction. The upper ($\tau=A$) and lower ($\tau=B$) chains are characterized by long-range exchange interactions within the chain,  $J^{\tau}_r$, where $r$ is the range of interaction. The exchange interactions are assumed to be positive (antiferromagnetic), isotropic, and  different in the two chains. Here, we indicate only the nearest-neighbor (yellow arrow) and the next-nearest-neighbor (orange arrow) interactions. The two chains are weakly coupled by an interchain interaction  $J_{\perp}$ (brown arrow) and subjected to a uniform magnetic field $b$ (blue arrow). (b) Kondo necklace model with long-range spin interactions.  Here, the spins belonging to chain $B$ are alternatingly displaced on the two sites of chain $A$. As a result, chain $B$ is split into two sub-chains with a doubled lattice constant, whose nearest-neighbor interaction $J_1^B$ is equivalent to the next-nearest-neighbor interaction $J_2^A$ of chain $A$.} 
	\label{fig:setup} 
\end{figure}

Recent ground-breaking experiments on magnetic adatoms placed on metal surfaces have paved the way to the controlled assembly of  individual spin-carrying units into coupled spin chains~\cite{Hirjibehedin06,Khajetoorians11,Menzel12,Bryant15,Farinacci18,Christensen16,Toskovic16,Ruby17,Kim18,Heinrich18,Kamlapure18,Choi19,Pawlak19,Yang19,Khajetoorians19,Liebhaber20,
Schneider20,Dinge21,Kuster21,Kuster21b,Schneider21,Kamlapure21,Jack21,Yang21}. While the main driving force behind this experimental effort has been the realization of Majorana bound states~\cite{Klinovaja13,Vazifeh11,NadiPerge13,Pientka13,Braunecker13}, the high level of control achieved in these designed chains allows us to envisage the possibility to explore even more exotic low-dimensional spin models with novel properties.

Motivated by these recent developments, we focus here on the spiral staircase Heisenberg ladder (SSHL)
~\cite{Brunger08,Aristov10}, which is described in terms of two weakly-coupled antiferromagnetic Heisenberg spin-$1/2$ chains with different exchange interactions, see Fig.~\ref{fig:setup} (a). 
A special case of the SSHL model is the SU(2) Kondo necklace model~\cite{Doniach77,Brunger08}, see Fig.~\ref{fig:setup} (b).
Using  bosonization methods~\cite{Giamarchi03} and density matrix renormalization group (DMRG) simulations~\cite{White92,Hauschild18},
we will show that these ladders can host  magnetic phases characterized by exotic  excitations with fractional spin values that can be simply tuned  by an external magnetic field. This fractionalization of spin is the analog of charge fractionalization in low-dimensional strongly interacting electron systems, the prime example being fractional quantum Hall phases~\cite{Arovas84,Jain90,Stormer99,Kane02,Teo14,Klinovaja14b,Laubscher21}. Moreover, we show that these fractional magnetic phases can be probed via the magnetization as well as by non-equilibrium spin currents~\cite{Meier03,Hoogdalem11,Nakata17,Du17}, 
 giving rise to fractional spin conductances quantized in universal values of $(g \mu_B)^2/h$,
 with $g$ the $g$-factor, $\mu_B$ the Bohr magneton,  and $h$ the Planck constant. In contrast to spin currents in itinerant systems~\cite{Culcer04}, spin transport in insulating magnets is extremely appealing for applications due to the absence of standard Joule heating~\cite{Trauzettel08,Takei14,Jungwirth18}. 
 
\textit{Model.} We focus on a system composed of two weakly coupled antiferromagnetic spin-$1/2$ chains, oriented along the $x$ direction and labeled by an index $\tau \in \{A, B\}$, with different exchange interaction strengths, see Fig.~\ref{fig:setup} (a). We describe each chain by a model Hamiltonian with long-range isotropic~\footnote{Our results can be straightforwardly generalized to anisotropic exchange interactions \cite{SM}} spin interactions 
\begin{align}
H_{0,\tau} &=  \sum_{i,r}J^{\tau}_{r}\left[\frac{1}{2}\left( S^+_{i+r,{\tau}}S^-_{i,{\tau}} + \text{H.c.}\right)+  S_{i+r,{\tau}}^z S^z_{i,\tau}\right].
\end{align}
Here, $\vec{S}_{i,\tau}$ is the spin-$1/2$ operator acting on the site $i$ of the $\tau$-th chain
and $J^{\tau}_{r}>0$ is the exchange coupling of range $r\geq 1$ in chain $\tau$.
In addition, an external uniform magnetic field $b$ is applied along the $z$ direction:  
\begin{equation}
H_{Z,\tau} =  -b\sum_{i} S_{i,\tau}^z,
\end{equation}
which allows one to control the magnetization of the chain. Here, we absorb the coupling constant $g\mu_B$ in $b$.
The  coupling between the two chains is also antiferromagnetic and given by
\begin{equation}
H_{{\rm inter}} = J_{\perp} \sum_{i}\left[\frac{1}{2}  \left(S^+_{i,A}S^-_{i,B}+\text{H.c.}\right)  + S_{i,A}^z S^z_{i,B}\right] \label{eq:ham_inter},
\end{equation}
where the exchange interaction $0<J_{\perp}\ll J^{\tau}_r$ describes the weak interchain coupling. 
The total Hamiltonian is given by $H=\sum_{\tau=A,B} \left[H_{0,\tau} +H_{Z,\tau}\right]+  H_{{\rm inter}}$ and the modeled setup  is illustrated in Fig.~\ref{fig:setup} (a).
This model is the extension of the SSHL model~\cite{Brunger08,Aristov10} to the case of intrachain long-range interactions.

\begin{figure}[t]
\centering
\includegraphics[width=\linewidth]{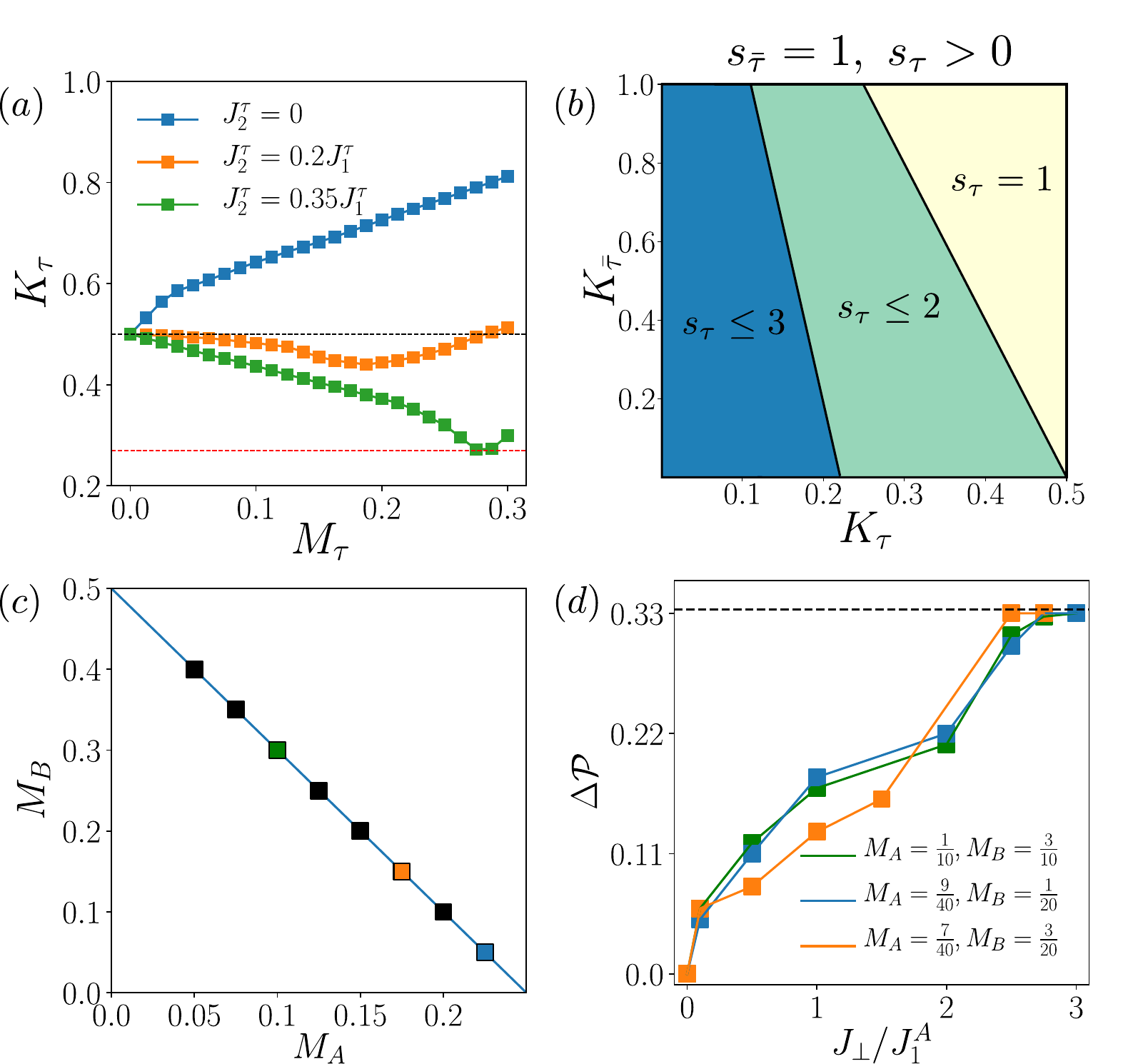}
\caption{(a) Luttinger liquid parameter $K_{\tau}$ computed by iDMRG simulations as function of magnetization $M_{\tau}$ for different values of the n.n.n.~interaction $J^{\tau}_r$. The presence of long-range interaction in a single chain is crucial in order to achieve desirable values of $K_{\tau}$ lower than $0.5$ (black dashed line). (b) Phase diagram for processes characterized by $(s_{\tau},s_{\bar{\tau}}=1)$ as  function of $K_{\tau}$ and $K_{\bar{\tau}}$ for  $s_{\tau}>0$. For $K_{\tau}>0.5$, no processes induce fractional excitations ($s_{\tau}=0$) except for require $|M_{\tau}|=|M_{\bar{\tau}}|$ (yellow).  For smaller values of $K_{\tau}$, fractional phases with $s_{\tau}=2$ (green) or $s_{\tau}=3$ (blue) are stabilized. (c) For magnetizations $M_A$ and $M_B$, which follow Eq.~\eqref{eq:resonance} for the process ($s_A=2$, $s_B=1$) (blue line),  we confirm numerically via DMRG (squares) fractional excitations $\Delta \mathcal{P}=1/3$  for strong interchain interaction. (d)  $\Delta \mathcal{P}$  is computed as  function of interchain coupling by DMRG at different values of magnetizations taken from panel (c): the colors of the lines correspond to the colors of the squares in panel (c). The interaction parameters are $J_2^A=0.5J_1^A$, $J_2^B=0.1J_1^A$ and $J_2^B=0.1J_1^A$.} 
	\label{fig:FigKM}
\end{figure}

\textit{Single chain.}  It is convenient to map each spin-$1/2$ chain onto a system of spinless fermions via Jordan-Wigner transformation: $S^z_{i,{\tau}} = c^{\dagger}_{i,{\tau}}c_{i,{\tau}}-\frac{1}{2}$ and
$S^{+}_{i,{\tau}} = c_{i,{\tau}}^{\dagger}\prod_{l<i} S^z_{l,{\tau}}\mathcal{T}_{i,\tau}$.~\cite{Galitski10,Hill17}.
Here, $c_{i,\tau}$ is a spinless fermionic annihilation operator acting on site $i$ of the $\tau$-th chain~\cite{Giamarchi03}. The product of spin operators is the standard Jordan-Wigner string that guarantees the proper anticommutation relation of fermions belonging to the same chain,  while $\mathcal{T}_{i,A/B}=\prod_{k<i}\tau_{k,A/B}^{y}\tau_{i,A/B}^x$ is a combination of Pauli matrices $\tau^{x,y}$  ensuring the correct fermion algebra for operators on different chains.

In fermionic representation, the magnetic field plays the role of a chemical potential. To treat the interaction between spinless fermions, it is useful to linearize the spectrum around the Fermi points. For this purpose, we can expand the fermionic operator as $c_{j,{\tau}} = e^{i k_{F,{\tau}} x_j} R_{{\tau}}(x_j) + e^{-i k_{F,{\tau}} x_j} L_{\tau}(x_j)$, where $x_j= j a$, with $a$ being  the lattice constant, and $R_{\tau}/L_{\tau}$ is the fermionic operator describing right/left-moving  fields.  Here, $k_{F,{\tau}} = \frac{\pi}{2a}\left[1+2 M_{\tau}\left(b/J^{\tau}_1\right)\right]$, where $M_{\tau}(b/J^{\tau}_1)$ is the $z$-component of the magnetization per site of chain $\tau$, defined as the expectation value of $S_{\tau}^z$ in the $\tau$-chain  in the absence of the interchain coupling. Note that the total magnetization in the SSHL is conserved.

To treat strong interactions, we employ bosonization techniques by introducing the conjugated bosonic fields $\phi_{\tau}$ and $\theta_{\tau}$ and the bosonization relation,
	\begin{equation}
	r_{\tau}\sim e^{-i \left(r \phi_{\tau}- \theta_{\tau}\right)},\label{bosonization}
	\end{equation}
where $r_{\tau} \in \{ R_{\tau}, L_{\tau}\}$ with correspondingly $r=\pm1$~\cite{vonDelft98,Giamarchi03}. 
In low-energy regime, each chain is described by a spinless Luttinger liquid (LL) characterized by an interaction parameter $K_{\tau}$, which depends on the strengths of exchange interactions of the microscopic model~\cite{Giamarchi03}. In the absence of magnetic fields and without long-range interaction, exact analytical expressions of $K_{\tau}$ can be derived via Bethe ansatz~\cite{Haldane80}. For the more general case considered here, we resort to the  infinite DMRG (iDMRG) algorithm, thus computing power-law correlation functions $\sum_{i}\left\langle S_{i,\tau}^{+}S_{i+j,\tau}^{-} \right\rangle$ and $\sum_{i}\left\langle S_{i,\tau}^{z}S_{i+j,\tau}^{z} \right\rangle$, from which we extract $K_{\tau}$~\cite{Giamarchi03,Ejima05,Ejima06}, see  Fig.~\ref{fig:FigKM} (a) 
 (see Supplemental Material (SM) for details~\cite{SM}).
In particular, next-nearest-neighbor (n.n.n.) couplings help to reach the  strong interaction regime where $K_{\tau}<0.5$.

\textit{Spiral staircase ladder - relevant perturbations.}  
Since the interchain coupling is assumed to be weak, one can treat it perturbatively. In the fermionic picture, the perturbations assume the form of multi-particle backscattering processes. 
 In order to avoid the problem of accounting for Jordan-Wigner strings~\cite{Galitski10,Hill17},
 we  consider only terms which do not transfer fermions (i.e. magnetization in spin space) between chains The most general perturbation one can construct reads~\cite{SM}
\begin{equation}
\mathcal{H}_{s_{\tau}s_{\bar{\tau}}} = g_{s_{\tau}s_{\bar{\tau}}}  e^{2i \left[s_{\tau}k_{F,\tau}+s_{\bar{\tau}}k_{F,\bar{\tau}}\right]x_j}\left(L_{\tau}^{\dagger}R_{\tau}\right)^{s_{\tau}}\left(L_{\bar{\tau}}^{\dagger}R_{\bar{\tau}}\right)^{s_{\bar{\tau}}} +\text{H.c.},\label{eq:proc}
\end{equation}
where $s_{\tau}$ are non-zero integers, whose absolute values represent the total number of fermions that are backscattered from left- to  right-moving channel in chain $\tau$. By using the relation between $k_{F,\tau}$ and  $M_{\tau}$, the resonance condition for the processes 
can be expressed as
\begin{equation}
s_{\tau} M_{\tau} + s_{\bar{\tau}}M_{\bar{\tau}} =p -  \frac{s_{\tau} +  s_{\bar{\tau}}}{2},\label{eq:resonance}
\end{equation}
where $p$ is integer. 

By applying Eq.~(\ref{bosonization}), the perturbation in Eq.~\eqref{eq:proc} assumes the bosonic form
\begin{eqnarray}
\mathcal{H}_{s_{\tau}s_{\bar{\tau}}} =\tilde{g}_{s_{\tau}s_{\bar{\tau}}} \int dx~\cos\left[2s_{\tau}\phi_{\tau}(x)+2s_{\bar{\tau}}\phi_{\bar{\tau}}(x)\right].\label{Eq:cos}
\end{eqnarray}
The next salient element of our analysis is the renormalization group (RG) flow of coupling constants. Since the coupling between the two chains is assumed to be weak, the scaling dimension for the interchain processes can be straightforwardly derived~\cite{vonDelft98,Giamarchi03} and  is given by $\Delta_{s_As_B} = s_A^2 K_{A} +s_B^2 K_{B}$. Importantly, the scaling dimension of processes transferring fermions between chains are always strictly larger than $\Delta_{s_As_B}$ for a fixed pair ($s_A,s_B$). Therefore, it is a good assumption to neglect them~\cite{SM}. In Fig.~\ref{fig:FigKM} (b), we  present the phase diagram for the term with $s_{\bar{\tau}}=1$ and $1 \le s_{\tau}\le 3$. In this diagram each region corresponds to values of LL parameters for which the processes with corresponding pair ($s_{\tau}$,$s_{\bar{\tau}}=1$) are relevant, i.e. with scaling dimensions less than two. If $s_{\tau}=0$,  the two chains are uncoupled. Here, we focus on $s_{\tau}\le3$ since for higher values of $s_{\tau}$, ever smaller 
$K_{\tau}$'s are required, which is hard to reach even with n.n.n. interactions (see Fig.~\ref{fig:FigKM}). We recall that while a perturbation can be relevant for certain  LL parameters according to the phase diagram in Fig.~\ref{fig:FigKM} (b), it can give rise to a term like Eq.~\eqref{Eq:cos} only if the resonant condition for the magnetization in Eq.~\eqref{eq:resonance} is simultaneously satisfied.

In the special case of equal magnetizations, $|M_{\tau}| = |M_{\bar{\tau}}|$,  the two symmetric processes ($s_{\tau},s_{\bar{\tau}}$) and  ($s_{\bar{\tau}},s_{\tau}$) can be stabilized simultaneously, thus giving rising to two commuting cosine perturbations. As a result, the spectrum becomes fully gapped. Here, we focus on the regime $|M_{\tau}|\ne |M_{\bar{\tau}}|$, in which the spin conductance can be finite. 
We emphasize that,  if the exchange interactions are different in the two chains, a uniform magnetic field suffices to induce different magnetizations.

\textit{Fractional spin excitations.} 
If RG relevant, the above perturbative processes result in the opening of a partial gap in the spectrum.  As striking consequence of this gap opening, fractional spin excitations  emerge in the system, as we show next. By using the  $z$-component of 
the total spin operator $S_z = (-1/\pi) \int dx \ \partial_x\left[\phi_A(x)+\phi_B(x)\right]$, one can compute the spin carried by an excitation created at the domain wall at which the argument of the cosine  in Eq.~\eqref{Eq:cos} jumps by $2\pi$~\cite{Chen17}. When $s_{\tau} = s_{\bar{\tau}}$, one finds that $\Delta S_z =
1/{s_{\tau}}$.
When $s_{\tau} \ne \pm s_{\bar{\tau}}$, it is convenient to change to a new bosonic basis,
\begin{equation}
\left(\begin{matrix}\theta^{(\tau)}_+\\ \phi^{(\tau)}_+\\\theta^{(\tau)}_-\\\phi^{(\tau)}_-\end{matrix}\right)  = 2 \left(\begin{matrix}s_{\tau}& 0 & -s_{\bar{\tau}} &  0 \\ 0 & s_{\tau} &0& s_{\bar{\tau}} \\ s_{\bar{\tau}}& 0 & -s_{\tau} &  0\\ 
0 & s_{\bar{\tau}} &0& s_{\tau}
\end{matrix}\right)\left(\begin{matrix}\theta_{\tau}\\\phi_{\tau}\\ \theta_{\bar{\tau}}\\ \phi_{\bar{\tau}} \end{matrix}\right),
\end{equation}
in which Eq.~(\ref{Eq:cos}) simplifies to $\mathcal{H}_{s_{\tau}s_{\bar{\tau}}} =\tilde{g}_{s_{\tau}s_{\bar{\tau}}} \int dx~\cos[\phi^{(\tau)}_+(x)]$.
The spin carried by the excitation created at the domain wall, where $\phi^{(\tau)}_+$ jumps by $2\pi$, is given by 
\begin{equation}
\Delta S_z =\frac{1}{s_{\tau}+s_{\bar{\tau}}}\frac{1}{2 \pi}\left[\phi^{(\tau)}_+(x)\right]^{+\epsilon}_{-\epsilon} = \frac{1}{s_{\tau}+s_{\bar{\tau}}}.\label{eq:frac_spin}
\end{equation}
 For $K_{\tau}>0.5$ only the process with $s_{\tau}=s_{\bar{\tau}}=1$ is RG relevant (see  Fig.~\ref{fig:FigKM} (b)), and, thus, $\Delta S_z=1$, i.e. the spin excitations  are standard magnons in this case. Since we are interested in processes for which fractional excitations can emerge, smaller values of $K_{\tau}$ are required, which is indeed achievable for strong n.n.n. interactions. 
In this case, spin excitations in the SSHL model  carry fractional spin given by $|\Delta S_z|$. This is one of our main results.

These fractional spin excitations can be tested numerically in a finite system of size $L$ by  introducing a generalized magnetic dipole  moment $\mathcal{P} =\frac{1}{L}\sum_{i=1}^{L}i \left(M_{\tau,i}+M_{\bar{\tau},i}\right)$, where $M_{\tau,i}$ is the average value of the out-of-plane magnetization at each site $i$. The difference  $\Delta  \mathcal{P}$ computed in the different ground states of the partially gapped sector, where $\phi^{(\tau)}_+$ jumps by $2\pi$, is equivalent to $\Delta S_z$ (see SM~\cite{SM}).
By means of finite DMRG simulations, we computed $\Delta \mathcal{P}$ for various sets of magnetizations and interactions stabilizing the processes ($s_A=2,s_B=1$) characterized by $\Delta S_z=1/3$. The values of magnetization [squares in Fig.~\ref{fig:FigKM} (c)] for which we find the fractional values $\Delta \mathcal{P}=1/3$ follow the resonance condition of Eq. \ref{eq:resonance}. Moreover, in Fig.~\ref{fig:FigKM} (d) we show $\Delta \mathcal{P}$ as a function of the interchain coupling $J_{\perp}$ for three pairs of magnetizations taken from Fig.~\ref{fig:FigKM} (c). Indeed, the expected fractional value of $\Delta \mathcal{P}$ emerges only in the strong coupling regime, which agrees with our RG flow arguments.

\textit{Fractional spin conductance.}
Next, we address the question how to probe such fractional spin excitations. One possibility is for instance given by the dynamical structure factor $\mathcal{S}(q,\omega)$. However, while this quantity does show a partial gap and depends on the quantum numbers $s_\tau$, it seems difficult to extract them from  $\mathcal{S}(q,\omega)$  (see~\cite{SM}). Another, more promising possibility, is to study the non-equilibrium spin transport~\cite{Meier03} for  finite-sized systems. To see this, we compute the spin conductance in a two-terminal configuration, where we assume that the magnetic field changes along the sample~\cite{Ronetti19}. More specifically, we divide an infinite spin ladder into three regions: one central region of length $L$ subjected to the magnetic field $b$ and two semi-infinite regions (right and left leads), defined by $|x|>L/2$, subjected to the magnetic fields $b^L+\Delta b/2$ and $b^L-\Delta b/2$, respectively. Note that, in general, $b^L \ne b$. In linear response, the spin conductance $G_S$  is defined via a spin current $I_S = G_S \Delta b/(g \mu_B)$, describing the flow of the $z$-component of spin by a field gradient~\cite{Nakata17}. 
The distance between chains in the leads is assumed to be larger than in the central region such that interchain processes are negligible in the lead region and each of the two spin chains  can be described as spinless LL.
The spin conductance $G_S$ can then be computed by standard methods~\cite{Meier03,Meng14b,Shavit19,Ronetti20} (see SM for details~\cite{SM}),
\begin{equation}
	G_S=\frac{(g \mu_B)^2}{h}\frac{K_A^L K_B^L \,(1-s_{AB})^2}{
		K_A^L s_{AB}^2+K_B^L },
	\end{equation}
where we restored Planck constant $h$ and introduced the ratio $s_{AB}=s_A/s_B$. Due to the presence of $K_{A/B}^L$, $G_S$ is not universal and depends on the magnetic field $b^L$ in the leads. However, a substantial simplification occurs in the case $K_A^L = K_B^L\equiv K^L$, where one finds $G_S/G_S^{0}=(1-s_{AB})^2/
		\left(1+s_{AB}^2\right)$, where $G_S^{0} = K^L (g \mu_B)^2/h$ is the spin conductance of one gapless spin chain 
		and which depends on the values of exchange interactions in the lead regions only via $K^L$. 
		Interestingly, when the magnetic field $b^L$ is so small that both chains have a vanishing magnetization in the leads, the corresponding 
		parameters become universal, $K_{A/B}^L=1/2$~\cite{Giamarchi03}, regardless of the strength of all exchange interactions (see Fig.~\ref{fig:FigKM}). 
		Remarkably, the spin conductance becomes now an entirely universal fraction of $(g \mu_B)^2/h$, explicitly given by
	\begin{align}
	{G_S}&= \frac{(g \mu_B)^2}{2h}  \frac{(1-s_{AB})^2}{1+s_{AB}^2}.
	\end{align} 
Importantly,  the ratio $s_{AB}$  also defines the slope in the resonance condition between the two magnetizations defined in Eq. \ref{eq:resonance}, see Fig.~\ref{fig:FigKM} (c).

\textit{Experimental feasibility.} 
To discuss the feasibility of our proposal, we focus here on the simplest case with $s_A = 1$ and $s_B = 2$, giving rise to excitations with fractional spin $1/3$,
coming from terms $\propto(J_{\perp})^3$ with scaling dimension $\Delta_{1,2} = K_{A} + 4K_{B}$. 
For this case to occur, the magnetization values  have to be tuned to $M_A\sim 0$ and $M_B \sim 1/4$, in order to satisfy the resonance condition in Eq.~\eqref{eq:resonance} and to induce  
the values  $K_A\sim0.5$ and $K_B\le 0.375$ in the presence of  strong enough n.n.n. interactions. These magnetization values can be achieved in the limit where one of the two chains has a much stronger nearest-neighbor exchange interaction than the other, say, $J_{1}^{A} \gg J_{1}^{B}$. In this case, chain A remains at $M_{A}\sim0$ for a small magnetic field $b\ll J_1^{A}$, while the magnetization of  chain  B is tunable to $M_{B}\sim1/4$ , since one can have $b\sim J_1^{B}$. The limit $J_{1}^{A} \gg J_{1}^{B}$ is naturally achieved in the Kondo necklace model~\cite{Doniach77,Kiselev05}, shown in Fig.~\ref{fig:setup} (b). 
In this setup, the spins in one of the two chains are alternatingly placed on  two different sites of the other chain: this effectively splits the chain into two sub-chains with a doubled lattice constant. In the presence of  full isotropy between the two chains, a new relationship between the exchange interactions is enforced: $J_1^B = J_2^A$. By assuming a fast decay between nearest-neighbor and n.n.n. interaction, one obtains the limit  $J_1^B=J_2^A \ll J_1^A$. If $J_2^B$ is still large enough, then also  values $K_B\le 0.375$, which are necessary for the fractional phase, can be reached. For other examples of fractional spin excitations, see SM~\cite{SM}.

The spin conductance for this process could be detected e.g.  by STMs~\cite{Khajetoorians11,Schneider21} or  NV-centers~\cite{Du17,VanderSar2020}. Tuning both magnetizations to keep  the value of the fractional spin conductance fixed, the chosen process stays in resonance and one can access $s_{AB}$ via the slope. Magnetizations can be varied in the Kondo necklace model by tuning the magnetic field and also the ratio $J_1^A/J_1^B$, which can be controlled by changing the lattice spacing of $A$ or $B$~\cite{Dinge21,Kuster21}. Thus, from the measurement of the spin conductance, the ratio $s_{AB}$ can be obtained in two different and independent ways, i.e by the value of fractional spin conductance itself and the slope of the resonance. If the outcomes agree, this would be striking evidence for the existence of the partially gapped fractional phases. Moreover, taking into account that $s_{AB}$  is a ratio between two integers that are not expected to be larger than three, we expect that knowing $s_{AB}$ allows one easily to find $s_A$ and $s_B$, and, thus, also to determine $\Delta S_z$ uniquely. 

\textit{Conclusions.} We  considered the SSHL with magnetic field and long-range exchange interactions. We have shown that strong n.n.n. interactions in each chain generate perturbative processes, characterized by two integers  $s_{A}$ and $s_B$. The induced phases have a partial gap with excitations carrying  fractional spin $1/|s_{A} +s_B|$. To probe such excitations we considered a spin transport setup.We found that the spin conductance becomes a material-independent universal fraction of $(g \mu_B)^2/h$, expressed in terms of the ratio $s_{A}/s_{B}$. Further, we showed that the same ratio can be accessed independently via the slope of the resonance condition for the tuned magnetizations. We expect that our results can be generalized to the ferromagnetic case including Dzyaloshinskii-Moriya interaction. Also, while  we focused here on Abelian fractional excitations, we  expect that our approach can be extended to the non-Abelian case as well, which would be more interesting for topological qubits. Finally, we believe that, although being challenging, the proposed setup can be engineered with magnetic adatoms assembled on surfaces of metals
 ~\cite{Choi19,Khajetoorians19}. 
 
\textit{Acknowledgments.} This work was supported by the Swiss National Science Foundation and NCCR QSIT. This project received funding from the European Union's Horizon 2020 research and innovation program (ERC Starting Grant, grant agreement No 757725).

\bibliography{spin_chains.bib}

\begin{thebibliography}{103}
\bibitem [{Wig()}]{WignerString}%
  \BibitemOpen
  \href@noop {} {\bibinfo  {journal} {Such terms appear at magnetizations
  different from those we focus on and require much stronger interactions (with
  larger scaling dimensions) than what is considered here}\ }\BibitemShut
  {NoStop}%
\end{thebibliography}
\begin{thebibliography}{9}
\bibitem{WignerString} Such terms appear at  magnetizations different from those we focus on and require much stronger  interactions (with larger scaling dimensions) than what is considered here.
\end{thebibliography}

\widetext
\begin{center}
	\textbf{\large Supplemental Material: Fractional spin excitations and conductance in the spiral staircase Heisenberg ladder}\\
	\vspace{8pt}
	Flavio Ronetti, 
	Daniel Loss, and Jelena Klinovaja\\ \vspace{4pt}
	{\it Department of Physics, University of Basel,
		Klingelbergstrasse 82, CH-4056 Basel, Switzerland}
\end{center}

\setcounter{section}{0}
\setcounter{equation}{0}
\setcounter{figure}{0}
\makeatletter
\renewcommand{\thesection}{S\arabic{section}}
\renewcommand{\theequation}{S\arabic{equation}}
\renewcommand{\thefigure}{S\arabic{figure}}
\titleformat{\section}[hang]{\large\bfseries}{\thesection.}{5pt}{}

\section{Perturbative treatment of the interchain coupling via bosonization in the SSHL model}
\subsection{Details of numerical calculation of Luttinger liquid parameters}
In this part we discuss the numerical method that we used to obtain the plot for the Luttinger liquid parameters $K_{\tau}$ in Fig.~2 of the main text. The correlation functions are computed by means of the infinite density matrix renormalization group (iDMRG) algorithm. The algorithm is implemented using the library TenPy based on tensor networks tools, such as matrix product states (MPS) and matrix product operators (MPO)~\cite{TenPy}. The simulation is run with a fixed magnetization: the unit cells of the MPS are chosen such that they can be commensurate with the specific magnetization. The final length of the systems is usually of between $10^{3}$ and $10^{4}$ sites. The maximum bond dimension is fixed to $\chi=1024$: with this value we reach a truncation error of the order $10^{-8}$ or less. 
\begin{figure}[h]
	\centering
	\includegraphics[width=0.9\linewidth]{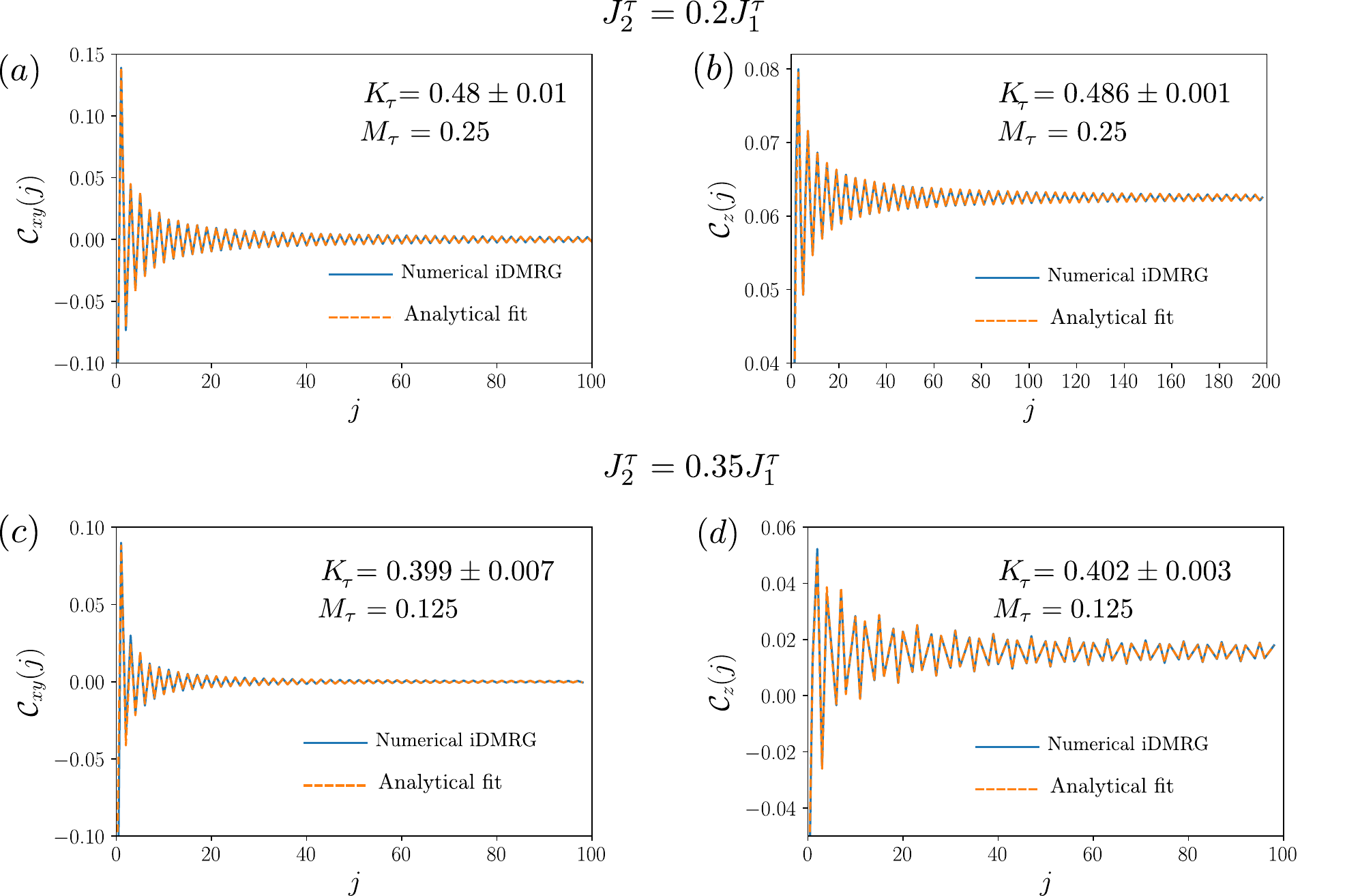}
	\caption{Correlation functions for the chain $\tau$ and corresponding fitted Luttinger liquid parameter  $K_{\tau}$ for a fixed magnetization $M_{\tau}$ and next-nearest-neighbor interaction strength $J_2^{\tau}$. (a) In-plane correlation function $\mathcal{C}_{xy}$. The parameters are $J_2^{\tau}=0.2J_1^{\tau}$ and $M^{\tau}=0.25$. (b) Out-of-plane correlation function $\mathcal{C}_{z}$. The parameters are $J_2^{\tau}=0.2J_1^{\tau}$ and $M^{\tau}=0.25$. (c) In-plane correlation function $\mathcal{C}_{xy}$. The parameters are $J_2^{\tau}=0.35J_1^{\tau}$ and $M^{\tau}=0.125$. (d) Out-of-plane correlation function $\mathcal{C}_{z}$. The parameters are $J_2^{\tau}=0.35J_1^{\tau}$ and $M^{\tau}=0.125$.  The choice of next-nearest-neighbor interaction strength $J_2^{\tau}$ corresponds to the one made for Fig. 2 in the main text.}
	\label{fig:iDMRG}
\end{figure}
The discretized version of the correlation functions is
\begin{align}
\mathcal{C}_{xy}(j) = \frac{1}{L}\sum_{n=1}^{L}\langle S^{+}_{n+j}S^{-}_n\rangle ,\\
\mathcal{C}_{z}(j) =\frac{1}{L}\sum_{n=1}^{L}\langle S^{z}_{n+j}S^{z}_n\rangle  ,
\end{align}
where $L$ is the maximum range at which the correlation function is computed. Typically, we choose $L=200$ sites. These correlation functions are fitted by using expressions obtained in the continuum limit for a spinless Luttinger liquid~\cite{Giamarchi}
\begin{align}
\mathcal{C}_{xy}(x) &= C_1 \frac{\cos(2\pi M_{\tau} x)}{x^{\frac{2K_{\tau}+1}{2 K_{\tau}}}} + C_2 \frac{\cos(\pi x)}{x^{\frac{1}{2K_{\tau}}}},\\
\mathcal{C}_{z}(x) &= M_{\tau}^2 - \frac{K_{\tau}}{2\pi^2 x^2} + C_3 \frac{\cos\left[\left(2M_{\tau}+1\right)\pi x\right]}{x^{2K_{\tau}}}.
\end{align}
This allows us to determine $K_{\tau}$ numerically, see Fig. \ref{fig:iDMRG}.

\subsection{Scattering processes for two coupled spin chains \label{sm:sec1}}
We consider the SSHL model consisting of two weakly coupled antiferromagnetic spin-$1/2$ chains, labeled as $A$ and $B$, described by the Hamiltonians $H=\sum_{\tau=A,B} \left[H_{0,\tau} +H_{Z,\tau}\right]+ H_{{\rm inter}}$, where the interchain coupling Hamiltonian is given by
\begin{equation}
H_{\rm inter} =  \sum_{j} \left[\frac{J^{xy}_{\perp}}{2}\left(S^+_{j,A}S^-_{j,B} + S^-_{j,A}S^+_{j,B}\right) + J^z_{\perp} S^z_{j,A}S^z_{j,B}\right].\label{SM:intra0}
\end{equation}
We assume that both $J^{xy}_{\perp}$and $J^z_{\perp}$ are positive and small such that we can treat $H_{\rm inter}$ as a perturbation. We note that in the main text we focused on the isotropic case $J^{xy}_{\perp}=J^z_{\perp}\equiv J_{\perp}$.
The spin operators can be expressed in terms of fermionic fields in the continuum limit:
\begin{align}
S^z_{j,A}S^z_{j,B}& \rightarrow \Big\{\frac{1}{4\pi^2}\partial_{x_j} \phi_A(x_j)\partial_{x_j}\phi_B(x_j) + e^{-i\left(2 k_{F,A} +2 k_{F,B}\right)x_j} R_A^{\dagger}(x_j)L_A(x_j)R_B^{\dagger}(x_j)L_B(x_j) \nonumber\\&+ e^{-i\left(2 k_{F,A} -2 k_{F,B}\right)x_j} R_A^{\dagger}(x_j)L_A(x_j)L_B^{\dagger}(x_j)R_B(x_j)+ \text{H.c.} + \dots \Big\},\label{eq:Szperp}\\
S^+_{j,A}S^-_{j,B}& \rightarrow e^{-i\left[\phi_{A}(x_j)-\phi_{B}(x_j)\right]}\Big[e^{i\left(k_{F,A}-k_{F,B}\right)x_j}R_A^{\dagger}(x_j)R_B(x_j) + e^{-i\left(k_{F,A}-k_{F,B}\right)x_j}L_A^{\dagger}(x_j)L_B(x_j) \nonumber\\&+ e^{-i\left(k_{F,A}+k_{F,B}\right)x_j}R_A^{\dagger}(x_j)L_B(x_j) + e^{i\left(k_{F,A}+k_{F,B}\right)x_j}L_A^{\dagger}(x_j)R_B(x_j)\Big],\label{eq:SpSm}
\end{align}
where $x_j = j a$, with $a$ being the lattice constant. The first term in Eq.~\eqref{eq:Szperp}, which includes density-density interactions between the chains, has to be added directly to the kinetic part of $H_{0,A} + H_{0,B}$. By using the remaining terms in $H_{\rm inter}$, one can construct perturbations in the fermionic picture. If we focus only on the operator part, they assume the generic form 
	\begin{equation}
	\left(R_A^{\dagger}L_AR_B^{\dagger}L_B\right)^{u_1}	\left(R_A^{\dagger}L_AL_B^{\dagger}R_B\right)^{u_2}\left(R_A^{\dagger}R_B\right)^{t_1}\left(L_A^{\dagger}L_B\right)^{t_2}\left(R_A^{\dagger}L_B\right)^{t_3}\left(L_A^{\dagger}R_B\right)^{t_4}e^{-i\left(t_1+t_2+t_3+t_4\right)\left(\phi_{A}-\phi_{B}\right)},
	\label{eq:SpSm2}
	\end{equation}
	where $u_{j}$ and $t_{j}$ are integers corresponding to the number of times each term from Eqs.~\eqref{eq:Szperp} and~\eqref{eq:SpSm} appears in the above expression Eq.~\eqref{eq:SpSm2}. When $u_{j}<0$ or $t_{j}<0$, one has to consider the hermitian conjugate of the terms in round brackets to the power $|u_{j}|$ or $|t_{j}|$, respectively. Here, all the operators are assumed to be evaluated at the same position along the $x$ axis. Let us now restore the dependence on Fermi momenta and on $x_j$:
	\begin{equation}
	e^{-i\Delta t\left[\phi_{A}(x_j)-\phi_{B}\right]}e^{i\left[\left(s_{R,A}-s_{L,A}\right)k_{F,A}+\left(s_{R,B}-s_{L,B}\right)k_{F,B}\right]x_j}R_A^{s_{R,A}}(x_j)L_A^{s_{L,A}}(x_j)L_B^{s_{L,B}}(x_j)R_B^{s_{R,B}}(x_j),\label{gen}
	\end{equation}
	where we introduced the integers
	\begin{align}
	s_{R,A} &= -\left(u_1+u_2+t_1+t_3\right),~~~s_{L,A} = u_1+u_2-t_2-t_4,\\
	s_{R,B} &= u_2-u_1+t_1+t_3,~~~s_{L,B} = u_1-u_2+t_2+t_4,\\
	\Delta t &= t_1+t_2+t_3+t_4.
	\end{align}
One can see that there exists the following constraint: $s_{R,A} + s_{R,B} + s_{L,A} + s_{L,B} = 0$. The latter is a consequence of the conservation of particle number. Momentum conservation imposes an additional constraint: 
\begin{equation}
\left(s_{R,A}-s_{L,A}\right)k_{F,A}+\left(s_{R,B}-s_{L,B}\right)k_{F,B} = \frac{2\pi}{a} p, \label{eq:consSM}
\end{equation}
where $p$ is an integer and where we accounted for the fact that in presence of a lattice the momentum is conserved only up to integer multiples of $2\pi / a$. In order to avoid the problem of having Jordan-Wigner string factors in the perturbative processes, we are considering only terms which do not transfer particles between chains~\cite{Galitski,Hill,WignerString}, i.e. we impose the additional constraint $\Delta t=0$. As a result, one has $s_{R,\tau}=-s_{L,\tau}$, where $\tau = A,B$. 
In spin language this means that there is no transfer of magnetization between the chains. Below, we will show that the processes with $\Delta t\ne 0$ always have a strictly larger scaling dimension than the perturbations that do not transfer particles between chains. By means of Eqs.~(\ref{eq:Szperp}) and~(\ref{eq:SpSm}), we can construct the following perturbation:
\begin{equation}
\mathcal{H}_{n_{\tau}m_{\tau}n_{\bar{\tau}}m_{\bar{\tau}}} = g_{n_{\tau}m_{\tau}n_{\bar{\tau}}m_{\bar{\tau}}}  e^{2i \left[(n_{\tau}+m_{\tau})k_{F,\tau}+(n_{\bar{\tau}}+m_{\bar{\tau}})k_{F,\bar{\tau}}\right]x_j}\left(L_{\tau}^{\dagger}R_{\tau}\right)^{m_{\tau}}\left(L_{\bar{\tau}}^{\dagger}R_{\bar{\tau}}\right)^{m_{\bar{\tau}}}\left(L_{\tau}^{\dagger}R_{\tau}\right)^{n_{\tau}}\left(L^{\dagger}_{\bar{\tau}}  R_{\bar{\tau}}\right)^{n_{\bar{\tau}}} +\text{H.c},\label{eq:procSM}
\end{equation}
where $m_{\tau}$,  $m_{\bar{\tau}}$, $n_{\tau}$, and $n_{\bar{\tau}}$ are integers with the constraint $(n_A+n_B) \mod 2 =0$, where $p \mod 2$ indicates the value of an integer $p$ modulo $2$. Here, we separated terms originating from  $J^{xy}_{\perp}$ and $J^{z}_{\perp}$ processes. 
In addition, if there are several ways to generate the same term in the perturbation expansion, we keep only the lowest order. As a result, we find
\begin{align}
g_{n_{\tau}m_{\tau}n_{\bar{\tau}}m_{\bar{\tau}}}\propto\left(J_{\perp}^{xy}\right)^{|m_{\tau}|+|m_{\bar{\tau}}|+ ||m_{\tau}|-|m_{\bar{\tau}}||}\left(J_{\perp}^{z}\right)^{\frac{|n_{\tau}+n_{\bar{\tau}}|}{2}+\frac{|n_{\tau}-n_{\bar{\tau}}|}{2}} =\left(J_{\perp}^{xy}\right)^{2\ {\max}\{[m_{\tau}|,|m_{\bar{\tau}}|\}}\left(J_{\perp}^{z}\right)^{\ {\max}\{[n_{\tau}|,|n_{\bar{\tau}}|\}}. \label{eq:procSM1}
\end{align}
As discussed in the main text, the properties of the system are solely determined by $s_{\tau}\equiv m_{\tau}+n_{\tau} \,\,  \in \mathbb{Z}$. This can be seen also directly from Eq. (\ref{eq:procSM}). In the simplest case, if $J_{\perp}^{xy}=0$ [$J_{\perp}^{z}=0$], we get $s_{\tau}\equiv n_{\tau}$ [$s_{\tau}\equiv m_{\tau}$]. If $J_{\perp}^{xy}$ and $J_{\perp}^{z}$ are of the same order of magnitude, one choose such a set of $m_\tau$ and $n_\tau$ that the order of the process in Eq. (\ref{eq:procSM1}) is minimized for a given pair $(s_{A}, s_B)$. If $(s_{A}+s_{B} )\mod 2 =0$, it is sufficient to include only $J_{\perp}^{z}$-processes by choosing $s_{\tau}\equiv n_{\tau}$. As a result, we arrive at the prefactor ${g}_{s_{\tau}s_{\bar{\tau}}} \propto (J_{\perp}^{z})^{\ {\max}\{[s_{\tau}|,|s_{\bar{\tau}}|\}}$. If $(s_{A}+s_{B} )\mod 2 =1$, one $J_{\perp}^{xy}$-process is to be included in Eq.~(\ref{eq:procSM1}).  Without loss of generality, we assume that $|s_{B}|>|s_{A}|$ and choose $n_A = s_A$ and $n_B= s_B-\text{sign}(s_{B})$. Hence, we arrive at the prefactor ${g}_{s_{A}s_{B}} \propto (J_{\perp}^{z})^{|s_B-\text{sign}(s_{B})|}(J_{\perp}^{xy})^2$. If $J_{\perp}^{xy}=J_{\perp}^{z}=J_{\perp}$, both cases can be summarized as ${g}_{s_{A}s_{B}} \propto  (J_{\perp})^{\max \{|s_A|,|s_B|\} + [(s_A+s_B) \mod 2]}$.

\subsection{Resonance condition for magnetization}
The processes previously introduced can be stabilized only for certain values of the magnetization of the two chains. Indeed, the oscillating exponential in Eq.~\eqref{gen} suppresses the perturbation when the Hamiltonian density is integrated over $x$. As a consequence, one has to fix the exponent of the complex exponential to an integer multiple of $2\pi i$. This condition, which in the fermion picture is equivalent to the conservation of momentum, can be expressed as
\begin{equation}
2 \left(s_{\tau}k_{F,\tau}+s_{\bar{\tau}}k_{F,\bar{\tau}}\right) = \frac{2\pi}{a}p.
\end{equation}
By using $k_{F,\tau} = \pi \left(1 + 2 M_{\tau}\right)/2a $, where $M_{\tau}$ is the magnetization (per site) in the chain $\tau$, one finds
\begin{equation}
s_{\tau}M_{\tau} + s_{\bar{\tau}}M_{\bar{\tau}} =p -  \frac{s_{\tau}+ s_{\bar{\tau}}}{2}.\label{SM:eq:mag1}
\end{equation}
Next, we note that $|M_{\tau}|\leq1/2$ by definition. However, the point $|M_{\tau}|=1/2$ corresponds to an empty band in fermion language where the  bosonization procedure is no longer valid and thus we only consider the case with $|M_{\tau}|<1/2$.  In this case, the possible values of $p$ are limited to $p_{min}< p < p_{max}$, where 
\begin{align}
p_{min} = \text{Min}\left\{s_{\tau}+s_{\bar{\tau}},\frac{2s_{\tau}+s_{\bar{\tau}}}{2},\frac{s_{\tau}+2s_{\bar{\tau}}}{2},\frac{s_{\tau}}{2},\frac{s_{\bar{\tau}}}{2},0\right\},&&&
p_{max} = \text{Max}\left\{s_{\tau}+s_{\bar{\tau}},\frac{2s_{\tau}+s_{\bar{\tau}}}{2},\frac{s_{\tau}+2s_{\bar{\tau}}}{2},\frac{s_{\tau}}{2},\frac{s_{\bar{\tau}}}{2},0\right\}.
\end{align}	
The condition in Eq.~\eqref{SM:eq:mag1} is the first constraint that the magnetizations of the two chains have to satisfy. The second one is related to the scaling dimension of the perturbations, which can be calculated by bosonizing the total Hamiltonian, as addressed next.

\subsection{Bosonized form of the Hamiltonian}
In order to bosonize the total Hamiltonian, it is a standard procedure to express the fermion fields as 
\begin{equation}
r_{{\tau}}\sim e^{-i \left(r\phi_{\tau}-\theta_{\tau}\right)},\label{eqSm:bosoId}
\end{equation}
where $r=\pm 1$ and $r_{\tau} \in \left\{R_{\tau},L_{\tau}\right\}$. The total Hamiltonian is rewritten as
\begin{equation}
H = \sum_{\tau=A,B} H_{0,\tau} + \sum_{\{s_{\tau},s_{\bar{\tau}}\}}\int dx~\mathcal{H}_{s_{\tau}s_{\bar{\tau}}}(x),
\end{equation}
where $\{s_{\tau},s_{\bar{\tau}}\}$ stands for all the possible perturbations that can be generated for every combination of integers $s_{\tau}$ and $s_{\bar{\tau}}$ that satisfies Eq.~\eqref{SM:eq:mag1} for fixed values of $M_{\tau}$ and $M_{\bar{\tau}}$. In the bosonized form, these contributions [see Eq.~(\ref{eq:procSM})]  can be rewritten as 
\begin{align}
&H_{0,\tau} = \int dx~\frac{u_{\tau}}{2\pi}\left\{\frac{1}{K_{\tau}}\left[\partial_x\phi_{\tau}(x)\right]^2+K_{\tau}\left[\partial_x\theta_{\tau}(x)\right]^2\right\},\\
&\mathcal{H}_{s_{\tau}s_{\bar{\tau}}}(x)= \tilde{g}_{s_{\tau}s_{\bar{\tau}}}\cos\left[2s_{\tau}\phi_{\tau}(x)+2s_{\bar{\tau}}\phi_{\bar{\tau}}(x)\right],  \label{perp}
\end{align} 
where we introduced the Luttinger liquid (LL) parameters $K_{\tau}$, whose specific values can be obtained numerically as shown in the main text (see Fig.~2 in the main text). The velocities $u_{\tau}$ depend on the magnetisation in the $\tau$-chain.
The coupling constant $\tilde{g}_{s_{\tau}s_{\bar{\tau}}}$ is proportional to ${g}_{s_{\tau}s_{\bar{\tau}}}$ in Eq.~(\ref{eq:procSM}) and has a dimension of energy density.
The scaling dimension can be directly obtained from Eq. (\ref{perp}) and it reads~\cite{Giamarchi}
\begin{equation}
\Delta_{s_{\tau}s_{\bar{\tau}}} = s_{\tau}^2 K_{\tau} + s_{\bar{\tau}}^2 K_{\bar{\tau}}.\label{SMeq:scaling}
\end{equation}
Since each LL parameter depends on the magnetization of the corresponding chain, the condition $\Delta_{s_{\tau}s_{\bar{\tau}}}<2$ is the second constraint that the magnetization has to satisfy. Moreover, we note that among all the possible perturbations with $\Delta_{s_{\tau}s_{\bar{\tau}}}<2$, only the most relevant one, namely the one with the smallest value of $\Delta_{s_{\tau}s_{\bar{\tau}}}$, is actually opening the gap. The other perturbations with a larger values of $\Delta_{s_{\tau}s_{\bar{\tau}}}$ do not contribute and are not taken into account in the effective Hamiltonian. As a result, one usually  works with a single cosine perturbation described by the pair of integers $(s_{\tau},s_{\bar{\tau}})$. In this case, the total Hamiltonian is
\begin{equation}
H = \sum_{\tau=A,B} H_{0,\tau} +\int dx~\mathcal{H}_{s_{\tau}s_{\bar{\tau}}}(x).
\end{equation}
The single cosine perturbation pins a combination of boson fields to a constant, thus opening a partial gap in the system. This partially gapped phases can host excitations carrying a fractional value of spin. In the remainder of this section, we relate the spin of the excitations to the pair of integerss $(s_{\tau},s_{\bar{\tau}})$.

\subsubsection{General case: $|s_{\tau}|\ne | s_{\bar{\tau}}|$}
Let us start by considering the case $|s_{\tau}|\ne | s_{\bar{\tau}}|$. The cosine perturbation $\mathcal{H}_{s_{\tau}s_{\bar{\tau}}}$ pins a linear combination of fields $\phi_{\tau}$ and $\phi_{\bar{\tau}}$. Therefore, it is convenient to choose a different basis of boson fields for which the argument of the cosine is given by a single boson field. To this purpose, let us introduce the transformation 
\begin{equation}
\left(\begin{matrix}\theta^{(\tau)}_+\\ \phi^{(\tau)}_+\\\theta^{(\tau)}_-\\\phi^{(\tau)}_-\end{matrix}\right)  = 2 \left(\begin{matrix}s_{\tau}& 0 & -s_{\bar{\tau}} &  0 \\ 0 & s_{\tau} &0& s_{\bar{\tau}} \\ s_{\bar{\tau}}& 0 & -s_{\tau} &  0\\ 
0 & s_{\bar{\tau}} &0& s_{\tau}
\end{matrix}\right)\left(\begin{matrix}\theta_{\tau}\\\phi_{\tau}\\ \theta_{\bar{\tau}}\\ \phi_{\bar{\tau}} \end{matrix}\right).\label{SM:eq:matrix}
\end{equation}
The commutation relations of new fields are defined as $\left[\phi^{(\tau)}_{\pm}(x),\theta^{(\tau)}_{\pm}(x')\right] = \pm 4\pi i \left(s_{\tau}^2-s_{\bar{\tau}}^2\right) \text{sign}(x-x')$. We note that the determinant of above transformation matrix is different from zero only for $s_{\tau}\ne \pm s_{\bar{\tau}}$. The cases $s_{\tau} = \pm s_{\bar{\tau}}$ will be considered below. In this new basis, the gap-opening term is rewritten as
\begin{equation}
\mathcal{H}_{s_{\tau}s_{\bar{\tau}}}(x)=\tilde{g}_{s_{\tau}s_{\bar{\tau}}}\cos\left[\phi^{(\tau)}_+(x)\right].
\end{equation}
In order to compute the values of spin for the excitations, it is useful to express the conserved $z$-component of the total magnetization 
\begin{equation}
S_z = -\frac{1}{\pi}\int dx \ \partial_x\left[\phi_A(x)+\phi_B(x)\right]\label{SM:eq:mag_dens0}
\end{equation}
in terms of these new operators as 
\begin{equation}
S_z =- \frac{1}{2\pi(s_{\tau}+s_{\bar{\tau}})}\sum_{l=\pm} \int dx~\partial_x\phi^{(\tau)}_l(x).\label{SM:eq:mag_dens}
\end{equation}
The total spin $\Delta S^{(\tau)}_z$ accumulated by an excitation around a kink (domain wall) at $x=0$ at which $\phi^{(\tau)}_+(x)$ jumps by $2\pi$  is given by 
\begin{align}
\Delta S^{(\tau)}_z &=- \frac{1}{2\pi(s_{\tau}+s_{\bar{\tau}})}\sum_{l=\pm} \int_{-\epsilon}^{\epsilon} dx~\partial_x\phi^{(\tau)}_l(x) = - \frac{1}{2\pi(s_{\tau}+s_{\bar{\tau}})} \left[\phi^{(\tau)}_+(x)\right]_{-\epsilon}^{\epsilon} =  \frac{2\pi}{2\pi(s_{\tau}+s_{\bar{\tau}})} = \frac{1}{s_{\tau}+s_{\bar{\tau}}}.\label{SM:eq:spin}
\end{align}
Again, $s_{\tau}$ and $s_{\bar{\tau}}$ are non-zero integers, which are also different in magnitude. Thus, we see that, generally, these excitations are characterized by a fractional value of spin $|\Delta S^{(\tau)}_z |=1/|{s_{\tau}+s_{\bar{\tau}}}|$. Finally, we  note  that the gapless mode $\phi^{(\tau)}_-$ is also associated with a fractional value of spin [see Eq.~\eqref{SM:eq:mag_dens}]. Below we show that this gapless mode is responsible for the fractional spin conductance.

\subsubsection{Special case: $s_{\tau}= \pm s_{\bar{\tau}}$}
Next, we focus on the special case $s_{\tau}= \pm s_{\bar{\tau}}$. The cosine perturbations can be rewritten as
\begin{align}
\mathcal{H}^{(\pm)}_{s_{\tau}}(x) =\tilde{g}^{(\pm)}_{s_{\tau}}\int dx~\cos\left\{2s_{\tau}\left[\phi_{\tau}(x)\pm \phi_{\bar{\tau}}(x)\right]\right\},
\end{align}
where we introduced the notation $\mathcal{H}^{(\pm)}_{s_{\tau}}\equiv\mathcal{H}_{s_{\tau} s_{\bar{\tau}}}$ and $\tilde{g}^{(\pm)}_{s_{\tau}}\equiv \tilde{g}_{s_{\tau}s_{\bar{\tau}}}$ for $s_{\bar{\tau}} = \pm s_{\tau}$, respectively. Let us comment that, in both cases, the cosine term pins a linear combination of boson fields which is independent of $s_{\tau}$. As a result, one can gather information about the excitations without changing the basis. We start by first considering $\mathcal{H}^{(-)}_{s_{\tau}}$. Here, the argument of the cosine does not affect the combination of the boson fields occurring in the total spin $S_z$ [see Eq.~\eqref{SM:eq:mag_dens0}]. As a consequence, there cannot exist excitations with a fractional value of spin. In the other case, the cosine perturbation directly pins the combination of fields which defines the magnetization density. Therefore, the total spin $\Delta S^{(\tau)}_z$ accumulated by an excitation around a kink at $x=0$ at which $2s_{\tau}\left[\phi_{\tau}(x)+ \phi_{\bar{\tau}}(x)\right]$ jumps by $2\pi$  is given by 
\begin{align}
\Delta S^{(\tau)}_z &=-\frac{1}{\pi} \int dx \hspace{1mm}\partial_x\left[\phi_A(x)+\phi_B(x)\right]=\nonumber\\& =-\frac{1}{2s_{\tau}}\frac{1}{ \pi}\left[2s_{\tau}\left\{\phi_{\tau}(x)+\phi_{\bar{\tau}}(x)\right]\right\}^{+\epsilon}_{-\epsilon} = \frac{1}{s_{\tau}}.
\end{align}
Also in this case, the spin is fractional if $s_{\tau}>1$. However, we note that the resonant processes with $s_{\tau}= s_{\bar{\tau}}>1$ occur for values of LL parameters much smaller than for processes with $s_{\tau}\ne s_{\bar{\tau}}$ with the same value of fractional spin excitations. Therefore, we focus in the main text only on the processes with $s_{\tau}\ne s_{\bar{\tau}}$.

\subsubsection{Example of a fractional state with spin $1/3$}

It is instructive to write down an explicit example of the leading perturbation and the corresponding term in spin representation. Let us focus on the case $s_{A}=1$ and $s_{B}=2$, which was also considered in the main text. This process satisfies the resonance condition, for instance, for $M_{A}\approx 0$ and $M_{B} \approx 1/4$. The fractional value of the spin for the excitations induced by this process is $\Delta S_z = 1/3$. The lowest order term in fermion representation is determined by $n_{A}=n_{B} = 1$, $m_{A}=0$, and $m_{B}=1$ and it can be written as
\begin{align}
\mathcal{H}_{s_{A}=1,s_{B}=2} &\propto (J^{xy}_{\perp})^2J^{z}_{\perp}
\left(L_{A}^{\dagger}R_{A}\right)
\left(L_{B}^{\dagger} R_{B}\right)^{2} +\text{H.c}.
\end{align}
The above perturbation comes from the following  contribution in spin representation:
\begin{align}
\mathcal{H}_{\rm pert} &\propto(J^{xy}_{\perp})^2J^{z}_{\perp} S_{j,A}^{+}S_{j,B}^{-}S_{j+1,B}^{+}S_{j+1,A}^{-}S^z_{j,A}S^z_{j,B}+\text{H.c}.
\end{align}  
The scaling dimension for this process is
\begin{equation}
\Delta_{1,2} = K_{A} + 4 K_{B}.
\end{equation}
For this term to be RG relevant and thereby opening a partial gap, we need $\Delta_{1,2} <2$, which can be satisfied for sufficiently small LL parameters $K_\tau$.
We note that there exist also other third order perturbations proportional to $(J^{xy}_{\perp})^3$, which carry a string. However, such terms can be safely ignored since this term has a larger scaling dimension compared to the one proportional to $(J^{xy}_{\perp})^2J^{z}_{\perp}$, due to the presence of the string factor $e^{-i\left[\phi_{A}(x_j)-\phi_{B}(x_j)\right]}$. Moreover, it would satisfy the resonant condition for magnetizations for a different pair of $M_{\tau}$ and $M_{\bar{\tau}}$, so that no competition between the terms  $\propto(J^{xy}_{\perp})^2J^{z}_{\perp}$  and the terms $\propto(J^{xy}_{\perp})^3$ can occur. Note that we have already excluded this kind of processes in Eq.~\eqref{eq:procSM}, since we allow only for perturbations proportional to an even power of $J^{xy}_{\perp}$.

\subsection{Two commuting perturbations}
For a very special choice of magnetization, it is possible to stabilize two commuting cosine terms. The resulting phase can be fully gapped and hosts two types of excitations. Let us consider two processes identified by the two pairs of  non-zero integers $(s^{(1)}_{\tau},s^{(1)}_{\bar{\tau}})$ and $(s^{(2)}_{\tau},s^{(2)}_{\bar{\tau}})$. Both perturbations have to be relevant in the RG sense, which implies the following constraints on the LL parameters [see Eq.~(\ref{SMeq:scaling})]:
\begin{equation}
\left(s^{(1)}_{\tau}\right)^2 K_{\tau} + \left(s^{(1)}_{\bar{\tau}}\right)^2 K_{\bar{\tau}}<2, ~~~\left(s^{(2)}_{\tau}\right)^2 K_{\tau} + \left(s^{(2)}_{\bar{\tau}}\right)^2 K_{\bar{\tau}}<2.
\end{equation}
Moreover, one also has to satisfy the following constraints on the magnetization, see Eq.~(\ref{SM:eq:mag1}), given by
\begin{align}
s^{(1)}_{\tau} M_{\tau} + s^{(1)}_{\bar{\tau}}  M_{\bar{\tau}} &=p_1 -  \frac{s^{(1)}_{\tau}+s^{(1)}_{\bar{\tau}}}{2},\label{SMeq:twocosine1}\\
s^{(2)}_{\tau}M_{\tau} + s^{(2)}_{\bar{\tau}}  M_{\bar{\tau}} &=p_2 -  \frac{s^{(2)}_{\tau}+s^{(2)}_{\bar{\tau}}}{2},\label{SMeq:twocosine2}
\end{align}	
which are solved by
\begin{align}
M_{\tau} = \frac{-2 p_1 s_{\bar{\tau}}^{(2)}+2 p_2 s_{\bar{\tau}}^{(1)}+s_{\tau}^{(1)} s_{\bar{\tau}}^{(2)}-s_{\tau}^{(2)}
	s_{\bar{\tau}}^{(1)}}{2 s_{\tau}^{(2)} s_{\bar{\tau}}^{(1)}-2 s_{\tau}^{(1)} s_{\bar{\tau}}^{(2)}}, ~~~
M_{\bar{\tau}} =\frac{-2 p_1 s_{\tau}^{(2)}+2 p_2 s_{\tau}^{(1)}-s_{\tau}^{(1)} s_{\bar{\tau}}^{(2)}+s_{\tau}^{(2)}
	s_{\bar{\tau}}^{(1)}}{2 s_{\tau}^{(1)} s_{\bar{\tau}}^{(2)}-2 s_{\tau}^{(2)} s_{\bar{\tau}}^{(2)}}.
\end{align}	
We note that this solution is valid only if $ s_{\tau}^{(1)} s_{\bar{\tau}}^{(2)}- s_{\tau}^{(2)} s_{\bar{\tau}}^{(2)}\ne 0$, which implies $s_{\tau}^{(1)}/s_{\tau}^{(2)}\ne s_{\bar{\tau}}^{(1)}/s_{\bar{\tau}}^{(2)}$. When $s_{\tau}^{(1)}/s_{\tau}^{(2)}= s_{\bar{\tau}}^{(1)}/s_{\bar{\tau}}^{(2)}$, the system is only partially gapped and we will comment on this trivial case separately. As an example, let us consider $(s^{(1)}_{\tau}=2,s^{(1)}_{\bar{\tau}}=1)$ and  $(s^{(2)}_{\tau}=-1,s^{(2)}_{\bar{\tau}}=2)$. The constraints on the magnetizations are satisfied for $M_{\tau} = -0.1$ and $M_{\bar{\tau}} = -0.3$ (if $p_1 = 1$ and $p_2 = 0 $). The scaling dimensions are both smaller than $2$ only if $K_{\tau}<0.2$ and $K_{\bar{\tau}}<0.2$ simultaneously. The latter condition can be achieved if n.n.n. interactions are strong enough or if we include  longer-range interaction than the n.n.n. one.

For the general case, $s_{\tau}^{(1)}/s_{\tau}^{(2)}\ne s_{\bar{\tau}}^{(1)}/s_{\bar{\tau}}^{(2)}$, the total Hamiltonian is rewritten as 
\begin{align}
H &= \int dx~\sum_{\tau = A,B}\frac{u_{\tau}}{2\pi}\left\{\frac{1}{K_{\tau}}\left[\partial_x\phi_{\tau}(x)\right]^2+K_{\tau}\left[\partial_x\theta_{\tau}(x)\right]^2\right\}\nonumber\\& + \tilde{g}_{s^{(1)}_{\tau}s^{(1)}_{\bar{\tau}}}\int dx~\cos\left[2s^{(1)}_{\tau}\phi_{\tau}(x)+2s^{(1)}_{\bar{\tau}}\phi_{\bar{\tau}}(x)\right]\nonumber\\& +  \tilde{g}_{s^{(2)}_{\tau}s^{(2)}_{\bar{\tau}}}\int dx~\cos\left[2s^{(2)}_{\tau}\phi_{\tau}(x)+2s^{(2)}_{\bar{\tau}}\phi_{\bar{\tau}}(x)\right], \label{re}
\end{align} 
where in the last two terms $\tau \in \{A,B\}$. The spectrum of the system is fully gapped. In the strong coupling regime, one can find the values to which $\phi_{\tau}(x)$ and $\phi_{\bar{\tau}}(x)$ are pinned. For this one has to satisfy the conditions
\begin{align}
2s^{(1)}_{\tau}\phi_{\tau}(x)+2s^{(1)}_{\bar{\tau}}\phi_{\bar{\tau}}(x) &= (2k_1+1)\pi,\\
2s^{(2)}_{\tau}\phi_{\tau}(x)+2s^{(2)}_{\bar{\tau}}\phi_{\bar{\tau}}(x) &= (2k_2+1)\pi,
\end{align}
giving the solutions 
\begin{align}
\phi_{\tau}(x)&= \frac{\pi  [-(2 k_1+1) s_{\bar{\tau}}^{(2)}+2 k_2 s_{\bar{\tau}}^{(1)}+s_{\bar{\tau}}^{(1)}]}{2s_{\tau}^{(2)}
	s_{\bar{\tau}}^{(1)}-2s_{\tau}^{(1)} s_{\bar{\tau}}^{(2)}},\\
\phi_{\bar{\tau}}(x) &=\frac{\pi  [-(2 k_1+1) s_{\tau}^{(2)}+2 k_2 s_{\tau}^{(1)}+s_{\tau}^{(1)}]}{2s_{\tau}^{(1)}
	s_{\bar{\tau}}^{(2)}-2s_{\tau}^{(2)} s_{\bar{\tau}}^{(1)}}.
\end{align}
Then, one can read off the value of the fractional spin carried by the excitations by using the magnetization  in Eq.~\eqref{SM:eq:mag_dens0}. Indeed, when the argument of the first (second) cosine perturbation jumps by $2\pi$, then one has that $k_1 \rightarrow k_1 - 1$ ($k_2 \rightarrow k_2 - 1$). As a result, the fractional spins associated with the modes $\pm$ are
\begin{align}
\Delta S^{+,\tau}_z = -\frac{s_{\tau}^{(2)}-s_{\bar{\tau}}^{(2)}}{s_{\tau}^{(1)} s_{\bar{\tau}}^{(2)}-s_{\tau}^{(2)} s_{\bar{\tau}}^{(1)}},~~~
\Delta S^{-,\tau}_z =- \frac{s_{\tau}^{(1)}-s_{\bar{\tau}}^{(1)}}{s_{\tau}^{(1)} s_{\bar{\tau}}^{(2)}-s_{\tau}^{(2)} s_{\bar{\tau}}^{(1)}}.
\end{align}
For the example considered above, one has $\Delta S^{+,\tau}_z =3/5$ and $\Delta S^{-,\tau}_z =1/5$. We point out that, in general, two types of excitations carrying a different value of fractional spin can co-exist, but this requires rather a high fine-tuning of the magnetizations.
We also remark that the special cases $s^{(1)}_{\tau}=s^{(1)}_{\bar{\tau}}$ or $s^{(2)}_{\tau}= s^{(2)}_{\bar{\tau}}$ are well defined as long as the condition $s_{\tau}^{(1)}/s_{\tau}^{(2)}\ne s_{\bar{\tau}}^{(1)}/s_{\bar{\tau}}^{(2)}$ still holds. In order to reproduce the case with a single cosine perturbation one has to choose $s^{(1)}_{\tau}=s^{(2)}_{\bar{\tau}}$ and $s^{(2)}_{\tau}=s^{(1)}_{\bar{\tau}}$, thus obtaining [see Eq.~\eqref{SM:eq:spin}]
\begin{align}
\Delta S^{+,\tau}_z = \Delta S^{-,\tau}_z=\frac{1}{s_{\tau}^{(1)} + s_{\bar{\tau}}^{(1)}}.
\end{align}

\subsubsection{Special case $s_{\tau}^{(1)}s_{\bar{\tau}}^{(2)}=s_{\tau}^{(2)}s_{\bar{\tau}}^{(1)}$}
Let us now consider the case when the condition $s_{\tau}^{(1)}s_{\bar{\tau}}^{(2)}=s_{\tau}^{(2)}s_{\bar{\tau}}^{(1)}$ is satisfied. In order to stabilize two perturbations, it is sufficient that one condition among~\eqref{SMeq:twocosine1} and~\eqref{SMeq:twocosine2} is satisfied, since the other one is then automatically valid too. For simplicity let us assume that $s_{\tau}^{(2)}/s_{\tau}^{(1)}\equiv\alpha > 1$. The Hamiltonian becomes
\begin{align}
H &= \int dx~\sum_{\tau = A,B}\frac{u_{\tau}}{2\pi}\left\{\frac{1}{K_{\tau}}\left[\partial_x\phi_{\tau}(x)\right]^2+K_{\tau}\left[\partial_x\theta_{\tau}(x)\right]^2\right\}\nonumber\\& + \tilde{g}_{s^{(1)}_{\tau}s^{(1)}_{\bar{\tau}}}\int dx~\cos\left[2s^{(1)}_{\tau}\phi_{\tau}(x)+2s^{(1)}_{\bar{\tau}}\phi_{\bar{\tau}}(x)\right]\nonumber\\& +  \tilde{g}_{s^{(2)}_{\tau}s^{(2)}_{\bar{\tau}}}\int dx~\cos\left\{\alpha\left[2s^{(1)}_{\tau}\phi_{\tau}(x)+2s^{(1)}_{\bar{\tau}}\phi_{\bar{\tau}}(x)\right]\right\}.
\end{align} 
In this case the system is only partially gapped because the second cosine potential is pinning exactly the same combination of boson fields as the first one. We note that it is not easy to have both these perturbations relevant in the RG sense. Indeed, the scaling dimension of process $(2)$ is $\alpha^2$ times bigger than the scaling dimension for the other processes: for instance, $K_{\tau}$ and $K_{\bar{\tau}}$ have to be $\alpha^2$ times smaller than in the case with only the first cosine perturbation. It is important to note that $\alpha$ gets closer to $1$ only when both $s_{\tau}^{(1)}$ and $s_{\tau}^{(2)}$ are very big: these processes can be relevant only for extremely small values of both LL parameters.

Let us now determine the fractional excitations. When $\alpha$ is an odd integer, $2s^{(1)}_{\tau}\phi_{\tau}(x)+2s^{(1)}_{\bar{\tau}}\phi_{\bar{\tau}}(x)$ is pinned around the minima of the first potential, since those are minima also of the second one. The value of fractional excitations is the same as the one obtained for a single cosine perturbation, as described previously in this section (see Eq.~\eqref{SM:eq:spin}).  When $\alpha$ is an even integer or a fraction, the value at which the combination $2s^{(1)}_{\tau}\phi_{\tau}(x)+2s^{(1)}_{\bar{\tau}}\phi_{\bar{\tau}}(x)$ is pinned depends on the values of the two amplitudes, since both cosine potentials must be minimized at the same time. Given the fact that the amplitude of the second processes is perturbatively smaller than the first one, one can expect that $2s^{(1)}_{\tau}\phi_{\tau}(x)+2s^{(1)}_{\bar{\tau}}\phi_{\bar{\tau}}(x)$ is still pinned mainly around the minima of the first cosine potentials. If the two amplitudes are comparable, two ground states can have a separation smaller than $2\pi$, depending on the values of the amplitudes of the processes: as a result, the value of fractional spins can be different from Eq.~\eqref{SM:eq:spin}. Finally, we note that above analysis for two cosine potentials can be generalized to any number of perturbations whose arguments are all integer multiples of $2s^{(1)}_{\tau}\phi_{\tau}(x)+2s^{(1)}_{\bar{\tau}}\phi_{\bar{\tau}}(x)$.

\begin{figure}[h]
	\centering
	\includegraphics[width=0.9\linewidth]{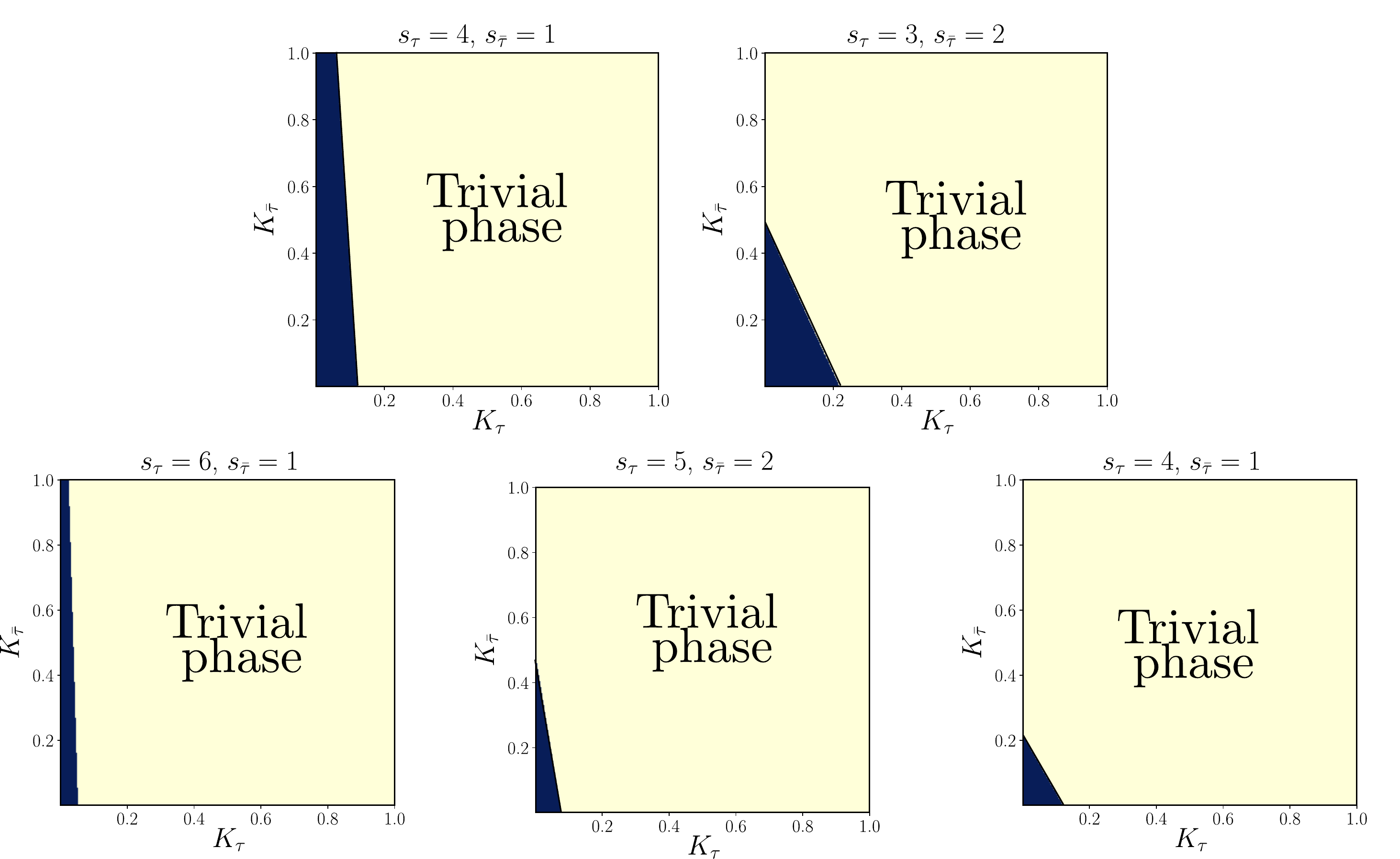}
	\caption{The phase diagram for the process $\mathcal{H}_{s_{\tau}s_{\bar{\tau}}}$  as a function of LL parameters of chains $\tau$ and $\bar{\tau}$ for different values of $s_{\tau} $ and $s_{\bar{\tau}} $.} 
	\label{fig:DiagramsSM}
\end{figure}

\subsubsection{Examples of other fractional states: scaling dimensions and phase diagrams}
Here, we discuss several cases in relation to their scaling dimension, in order to understand which processes are the most interesting ones. By using standard RG methods, one can show that the scaling dimension is given by Eq.~\eqref{SMeq:scaling}. Let us disregard processes with $s_{\tau}^2=s_{\bar{\tau}}^2$. Moreover, we are not interested in processes with $s_{\tau}=0$ or $s_{\bar{\tau}}=0$. As a consequence, it is clear that the processes with the smallest scaling dimension are those with ($s_{\tau} = 2$,$s_{\bar{\tau}} = 1$) or ($s_{\tau} = 1$,$s_{\bar{\tau}} = 2$). One can also easily see that, among processes inducing fractional spins with a bigger denominator, the ones with ($s_{\tau} = 3$,$s_{\bar{\tau}} = 1$) or ($s_{\tau} = 1$,$s_{\bar{\tau}} = 3$) are the next ones in terms of scaling dimensions. For higher values of ($s_{\tau} $,$s_{\bar{\tau}}$), the situation is more complicated and we plot some examples in Fig.~\ref{fig:DiagramsSM}. One can see that, the higher the values of $s_{\tau}$ ($s_{\bar{\tau}}$) the smaller $K_{\tau}$ ($K_{\bar{\tau}}$) has to be. For a fixed value of $s_{\tau} + s_{\bar{\tau}}$, there is no process which can dominate universally over the others. For instance, in the case of $s_{\tau} + s_{\bar{\tau}} = 5$ (first row of Fig.~\ref{fig:DiagramsSM}), when $s_{\tau}=4$ and $s_{\bar{\tau}}=1$, one needs smaller values of $K_A$ compared to the case $s_{\tau}=3$ and $s_{\bar{\tau}}=2$, but bigger values of $K_B$  are allowed than for $s_{\tau}=3$ and $s_{\bar{\tau}}=2$.

\subsection{Scaling dimension for processes with string factors}
	By using the bosonization identity in Eq.~\eqref{eqSm:bosoId}, one can show that the bosonized form of perturbations with $\Delta t \ne 0$ is given by
	\begin{equation}
	\mathcal{H}_{s_{\tau}s_{\bar{\tau}}}(x)\propto\cos\left\{2s_{\tau}\phi_{\tau}(x)+2s_{\bar{\tau}}\phi_{\bar{\tau}}(x)+\Delta t \left[\theta_{\tau}(x)-\theta_{\bar{\tau}}(x)\right]\right\}, 
	\end{equation} 
	Here, we omitted the coupling constant, since it is not necessary for the following discussion. The scaling dimension can be directly obtained from the above and it reads~\cite{Giamarchi}:
	\begin{equation}
	\Delta_{s_{\tau}s_{\bar{\tau}}\Delta t} = s_{\tau}^2 K_{\tau} + s_{\bar{\tau}}^2 K_{\bar{\tau}} + \frac{\Delta t^2}{4}\left(\frac{1}{K_{\tau}}+\frac{1}{K_{\bar{\tau}}}\right).\label{SMeq:scaling2}
	\end{equation}
	We note that the scaling dimension in Eq.~\eqref{SMeq:scaling} corresponds to the case $\Delta t =0$. Whenever $\Delta t \ne 0$, the scaling dimension of processes that do not transfer particles between chains, i.e the one in Eq.~\eqref{SMeq:scaling}, is always strictly smaller than $\Delta_{s_{\tau}s_{\bar{\tau}}\Delta t}$ in Eq. ~\eqref{SMeq:scaling}. As a consequence, the perturbations we considered in the main text are always more relevant in the RG sense than the processes with string factors, and that is why we can assume to disregard the terms with strings. Moreover, whenever $s_{\tau}s_{\bar{\tau}}\ne 0$, we note that the perturbations with $\Delta t\ne0$ are \textit{always irrelevant} in the regime of Luttinger liquid parameters we are interested in, i.e, $0<K_{\tau}<1$ and $0<K_{\bar{\tau}}<1$. Therefore, we conclude that it is completely justified to ignore the processes with $\Delta t \ne 0$ in the discussion in the main text.

\subsection{DMRG calculations for the fractional spins}
	\subsubsection{Indicator of fractional spins}
	Here, we show that by using DMRG one can obtain numerical evidence for the presence of excitations with fractional spin in the system. We focus on a finite size system of length $L$. The quantity that we compute with finite DMRG is 
	\begin{equation}
	\mathcal{P} =\frac{1}{L}\sum_{i=1}^{L}i \left(M_{\tau,i}+M_{\bar{\tau},i}\right),\label{eq:DMRGfrac}
	\end{equation}
	where $M_{\tau,i}$ is the average value of out-of-plane magnetization in each site $i$. This quantity can be considered as a generalized magnetic dipole moment , in analogy to the electric dipole moment or electric polarization which has proven to be useful for calculating fractional charges~\cite{Klara_numerics}.
	Below, by using bosonization techniques, we show that the difference in $\mathcal{P}$ (denoted by $\Delta\mathcal{P}$) computed in the different ground states of the gapped sector, where $\phi^{(\tau)}_+$ jumps by $2\pi$, is equivalent to $\Delta S_z$. The expression in terms of boson operators is rewritten as
	\begin{eqnarray}
	\mathcal{P} = - \frac{1}{2\pi L(s_{\tau}+s_{\bar{\tau}})}\sum_{l=\pm} \int_0^L dx~x\partial_x\phi^{(\tau)}_l(x) = - \frac{1}{2\pi L(s_{\tau}+s_{\bar{\tau}})}\sum_{l=\pm}\left[L \phi^{(\tau)}_l(L) -\int_0^L dx~\phi^{(\tau)}_l(x)\right]
	\end{eqnarray}
	where we made use of integration by parts. Each integral can be separated in a boundary contribution and a bulk contribution as
	\begin{equation}
	\int _0^Ldx~\phi^{(\tau)}_l(x) = \int_\epsilon^{L-\epsilon} dx~\phi^{(\tau)}_l(x) + \int_0^\epsilon dx~\phi^{(\tau)}_l(x) + \int_{L-\epsilon}^L dx~\phi^{(\tau)}_l(x),
	\end{equation}
	where $\epsilon$ is an arbitrarily small number. Then, one has 
	\begin{eqnarray}
	\mathcal{P} = - \frac{1}{2\pi L(s_{\tau}+s_{\bar{\tau}})}\sum_{l=\pm} \left[L \phi^{(\tau)}_l(L)-\int_\epsilon^{L-\epsilon} dx~\phi^{(\tau)}_l(x) - \int_0^\epsilon dx~\phi^{(\tau)}_l(x) - \int_{L-\epsilon}^L~\phi^{(\tau)}_l(x)\right].
	\end{eqnarray}
	Within the bulk, it is justified to replace $\phi^{(\tau)}_+$ with $\tilde{\phi}^{(\tau)}_{+}$, which is the value to which the field is pinned by the perturbation. Then, one can write
	\begin{eqnarray}
	\mathcal{P} = - \frac{1}{2\pi L(s_{\tau}+s_{\bar{\tau}})}\left\{\sum_{l=\pm} \left[L \phi^{(\tau)}_l(L) - \int_0^\epsilon dx~\phi^{(\tau)}_l(x) - \int_{L-\epsilon}^L~\phi^{(\tau)}_l(x)\right]-L \tilde{\phi}^{(\tau)}_+-\int_\epsilon^{L-\epsilon} dx~\phi^{(\tau)}_-(x)\right\}.
	\end{eqnarray}
	Next, we note that all quantities in the above expression do not change for different ground states \textit{except for} $\tilde{\phi}^{(\tau)}_+$.
	Since $\tilde{\phi}^{(\tau)}_+=2\pi n$, where $n=1,\dots,s_{\tau}+s_{\bar{\tau}}-1$, one can compute the difference of polarization in two degenerate ground states (say, for $n=1$ and $n=2$)
	\begin{eqnarray}
	\Delta \mathcal {P} =\mathcal{P} (n=2)-\mathcal{P} (n=1) =  \frac{2\pi}{2\pi (s_{\tau}+s_{\bar{\tau}})}  =\frac{1}{s_{\tau}+s_{\bar{\tau}}} = \Delta S_z.
	\end{eqnarray}
	By computing  $\mathcal {P}$ in different ground states, one can obtain numerical evidence for the presence of fractional spins in the system.
	\subsubsection{Details of DMRG simulations}
	In this part, we discuss the numerical method that we used to compute $\Delta \mathcal {P}$ in Fig.~2 of the main text. The correlation functions are computed by means of the finite density matrix renormalization group (DMRG) algorithm. The algorithm is implemented using the library TenPy based on tensor networks tools, such as matrix product states (MPS) and matrix product operators (MPO)~\cite{TenPy}. The simulation is run with fixed magnetizations for both chains: the lengths of the MPS are chosen such that they can be commensurate with both magnetizations. The lengths that we considered are $L=40$ and $L=50$. The maximum bond dimension is fixed to $\chi=64$: with this value we reach a truncation error of the order $10^{-6}$ or less.\\
	In order to obtain $\Delta \mathcal {P}$ we compute the ground states and several low-energy excited states: usually, we compute around four states. The excited states are obtained by finding the lowest energy states which are orthogonal to the previously found ground state and lower energy excited states.

To perform the calculations at fixed magnetizations, we consider the following form for the interchain Hamiltonian:
	\begin{equation}
	H_{\rm inter} =  \sum_{j} \left[\frac{J^{xy}_{\perp}}{2}\left(S^+_{j,A}S^-_{j,B} S^+_{j+1,B}S^-_{j+1,A} + \text{h.c.}\right) + J^z_{\perp} S^z_{j,A}S^z_{j,B}\right].\label{SMM:intra}
	\end{equation}
	We note that this Hamiltonian conserves magnetization in both chains separately. This choice is motivated by the fact that simulations with conserved magnetization are more efficient and fast. The processes that transfer magnetization between chains are excluded by construction, but this is not a crude approximation: we have shown that these interchain processes (involving strings, see above) are irrelevant in the RG sense and, therefore, we expect that they should not contribute in the strong coupling regime simulated by the lattice Hamiltonian in Eq.~\eqref{SMM:intra}. Moreover, we checked for some pairs of magnetizations that, even using the original microscopic Hamiltonian defined in Eq.~\eqref{SM:intra0}, $\Delta \mathcal{P}$ still assumes the expected fractional spin values. For the latter simulation, since we could not fix the magnetizations by a numerical constraint, we fixed the magnetic field in order to keep fixed the desired magnetizations in the chains. In our work, we did not check for effects of disorder but based on similar studies \cite{Klara_numerics}, we expect that fractional excitations are stable also against random disorder.

\section{Conductance for different LL  parameters and velocities} \label{sm:sec3}
\begin{figure}[t]
	\centering
	\includegraphics[width=0.7\linewidth]{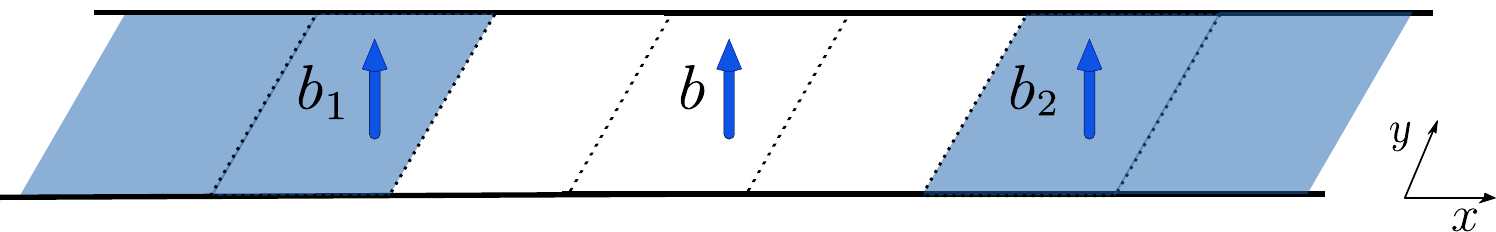}
	\caption{Setup to measure the fractional spin conductance. In the left and right leads (blue regions) different  magnetic fields ($b^1 = b^L + \Delta b/2$ and $b^2 = b^L - \Delta b/2$) are applied.} 
	\label{fig:conductance}
\end{figure}

In this Section we compute the conductance of two chains in presence of the backscattering process introduced in Eq.~\eqref{perp}. For definitness we fix $\tau = A$; the case with $\tau = B$ is completely identical and gives the same final result for the spin conductance. We recall the bosonized form for the perturbation 
\begin{equation}
\mathcal{H}_{s_{A}s_{B}}(x)=\tilde{g}_{s_{A}s_{B}} \cos\left[2s_{A}\phi_{A}(x)+2s_{B}\phi_{B}(x)\right].
\end{equation}
It is convenient to use the following change of basis
\begin{equation}
\left(\begin{matrix}
\theta_{\rho}(x) \\ \phi_{\rho}(x)\\\theta_{\sigma}(x)\\\phi_{\sigma}(x) 
\end{matrix}\right) = \frac{1}{\sqrt{2}}\left(\begin{matrix}
1 & 0 & 1 & 0 \\ 0 & 1 & 0 & 1 \\ 1 & 0 & -1 & 0 \\ 0 & 1 & 0 & -1
\end{matrix}\right) \left(\begin{matrix}
\theta_{A}(x) \\ \phi_{A}(x)\\\theta_{B}(x)\\\phi_{B}(x) 
\end{matrix}\right).\label{eq:rotation_matrix}
\end{equation}
Now, the process reads
\begin{equation}
\mathcal{H}_{s_{A}s_{B}}(x)=\tilde{g}_{s_{A}s_{B}} \cos\left\{\sqrt{2}\left[(s_A+s_B) \phi_{\rho}(x) +(s_A-s_B) \phi_{\sigma}(x)\right]\right\}.
\end{equation}
Let us assume that this perturbation is relevant and opens a gap. Due to the gap opening, the cosine potentials can be expanded to quadratic order in fields~\cite{Meng,Ronetti}, 
\begin{align}
\mathcal{H}_{s_{A}s_{B}}(x)&=\tilde{g}_{s_{A}s_{B}}\cos\left\{\sqrt{2}\left[c_{\phi_{\rho}} \phi_{\rho} +c_{\phi_{\sigma}} \phi_{\sigma}\right]\right\} = \\ &\sim \frac{\tilde{g}_{s_{A}s_{B}}}{2} \left\{\sqrt{2}\left[c_{\phi_{\rho}} \phi_{\rho} +c_{\phi_{\sigma}} \phi_{\sigma}\right]\right\}^2 = \tilde{g}_{s_{A}s_{B}}\left( c_{\phi_{\rho}}^2\phi_{\rho}^2 + 2c_{\phi_{\rho}}c_{\phi_{\sigma}} \phi_{\rho}\phi_{\sigma} +c_{\phi_{\sigma}}^2 \phi_{\sigma}^2\right).
\end{align}
where we introduced $c_{\phi_{\rho}} = s_A+s_B$ and  $c_{\phi_{\sigma}} =  s_A-s_B$.\\
After expanding to quadratic order the cosine potential and integrating out $\theta_{\rho}$ and $\theta_{\sigma}$, the action reads
\begin{equation}
\mathcal{S} =\frac{1}{2\pi} \int dx\int d\tau ~ \Psi(x,\tau)^{T} \mathcal{M}(x,\tau) \Psi(x,\tau),
\end{equation} 
where $\Psi(x,\tau)^T = \left(\phi_{\rho}(x,\tau),\phi_{\sigma}(x,\tau)\right)$ and
\begin{equation}
\mathcal{M}(x,\tau) = \left(\begin{matrix}\mathcal{M}_{11}(x,\tau)  & \mathcal{M}_{12}(x,\tau)\\ \mathcal{M}_{12}(x,\tau)& \mathcal{M}_{22}(x,\tau) \end{matrix}\right),\label{eq:action}
\end{equation}
with
\begin{align}
\mathcal{M}_{11}(x,\tau)&=-\frac{1}{2}\partial_x \left[\frac{u_{A}(x)}{K_{A}(x)}+\frac{u_{B}(x)}{K_{B}(x)}\right]\partial_x + \frac{1}{2} \left[\frac{1}{u_{A}(x)K_{A}(x)}+\frac{1}{u_{B}(x)K_{B}(x)}\right]\partial^2_\tau +  \tilde{g}_{s_{A}s_{B}}(x) c_{\phi_{\rho}}^2,\\
\mathcal{M}_{12}(x,\tau)&=-\frac{1}{2}\partial_x \left[\frac{u_{A}(x)}{K_{A}(x)}-\frac{u_{B}(x)}{K_{B}(x)}\right]\partial_x + \frac{1}{2} \left[\frac{1}{u_{A}(x)K_{A}(x)}-\frac{1}{u_{B}(x)K_{B}(x)}\right]\partial^2_\tau +  \tilde{g}_{s_{A}s_{B}}(x) c_{\phi_{\rho}}c_{\phi_{\sigma}},\\
\mathcal{M}_{22}(x,\tau)&=-\frac{1}{2}\partial_x \left[\frac{u_{A}(x)}{K_{A}(x)}+\frac{u_{B}(x)}{K_{B}(x)}\right]\partial_x + \frac{1}{2} \left[\frac{1}{u_{A}(x)K_{A}(x)}+\frac{1}{u_{B}(x)K_{B}(x)}\right]\partial^2_\tau +  \tilde{g}_{s_{A}s_{B}}(x) c_{\phi_{\sigma}}^2.
\end{align}
Here, velocities, coupling constant $\tilde{g}_{s_{A}s_{B}}$, and Luttinger liquid parameters are inhomogeneous quantities, they read:
\begin{align}
u_{A /B}(x) &= \begin{cases}
u_{A/B}^L &|x| > \frac{L}{2}\\u_{A / B} &|x| < \frac{L}{2}
\end{cases},~~~
\tilde{g}_{s_{A}s_{B}}(x) = \begin{cases}
0&|x| > \frac{L}{2}\\\tilde{g}_{s_{A}s_{B}} &|x| < \frac{L}{2}
\end{cases},~~~
K_{A/ B}(x) = \begin{cases}
K_{A/B}^L &|x| > \frac{L}{2}\\K_{A/ B} &|x| < \frac{L}{2}
\end{cases},
\end{align}
where $u_{A/B}^L $ and $K_{A/B}^L$ are the velocities and the Luttinger liquid parameter  inside the leads. The parameters in the leads and in the system are different since the magnetic field is non-uniform: $b^1 = b^L + \Delta b/2$ and $b^2 = b^L - \Delta b/2$, as shown in Fig.~\ref{fig:conductance}. Here, $\Delta b$ is a small detuning from the value $b^L$. In the framework of linear response theory, the spin current is given by $I_S = G_S \Delta b / g\mu_B$.

The Kubo formula for the two-terminal spin conductance is given by~\cite{Kane,Meng} 
\begin{equation}
\label{eq_condKubo}
G_S= 2(g \mu_B)^2\frac{\omega_n}{\pi^2} G_{cc}(x,\omega_n)\Big|_{i\omega_n \rightarrow \omega + i 0^+, \omega \rightarrow 0},
\end{equation}
where $G_{cc}$ is the Green function for $\phi_{\rho}$. In order to compute this conductance we have to solve for $G_{cc}$ the following set of differential equations~\cite{Meng} 
\begin{equation}
\mathcal{M}(x,\tau) \mathcal{G}(x,\tau) = \pi \delta\left(x-x'\right) \mathbb{I}_2,
\end{equation}
where $\mathcal{G}(x,\tau)$ is the matrix whose elements are all possible Green functions. In particular, we are interested in the four equations for $x\ne x'$, which involve $G_{cc}$.

To solve the system of differential equations inside the system (white region of Fig.~\ref{fig:conductance}), it is convenient to use the ansatz $G_{ij} \sim \epsilon_{ij}e^{q x}$ and move to Fourier space for the time variable. One finds
\begin{align}
&\tilde{M}\left(q,\omega_n\right) \left(\begin{matrix} \epsilon_{cc} \\ \epsilon_{cs}\end{matrix}\right) = \left(\begin{matrix} 0 \\ 0\end{matrix}\right),\label{eq:system}\\
&\tilde{M}\left(q,\omega_n\right) = \left(\begin{matrix}\tilde{M}_{11}\left(q,\omega_n\right) &\tilde{M}_{12}\left(q,\omega_n\right)\\\tilde{M}_{12}\left(q,\omega_n\right) &\tilde{M}_{22}\left(q,\omega_n\right) \end{matrix}\right),\\
\tilde{M}_{11}(q,\omega)&=-\frac{1}{2}\left[\frac{u_{A}}{K_{A}}+\frac{u_{B}}{K_{B}}\right]q^2 + \frac{1}{2} \left[\frac{1}{u_{A}K_{A}}+\frac{1}{u_{B}K_{B}}\right]\omega^2 +  \tilde{g}_{s_{A}s_{B}}c_{\phi_{\rho}}^2,\\
\tilde{M}_{12}(q,\omega)&=-\frac{1}{2}\left[\frac{u_{A}}{K_{A}}-\frac{u_{B}}{K_{B}}\right]q^2 + \frac{1}{2} \left[\frac{1}{u_{A}K_{A}}-\frac{1}{u_{B}K_{B}}\right]\omega^2 +  \tilde{g}_{s_{A}s_{B}} c_{\phi_{\rho}}c_{\phi_{\sigma}},\\
\tilde{M}_{22}(q,\omega)&=-\frac{1}{2}\left[\frac{u_{A}}{K_{A}}+\frac{u_{B}}{K_{B}}\right]q^2 + \frac{1}{2} \left[\frac{1}{u_{A}K_{A}}+\frac{1}{u_{B}K_{B}}\right]\omega^2 +  \tilde{g}_{s_{A}s_{B}} c_{\phi_{\sigma}}^2.
\end{align}
The values of $q$ can be determined by imposing that $\text{Det}\left(\tilde{M}\right)=0$ and are given by 
\begin{align}
q_1 &= - q_2 =\tilde{g}_{s_{A}s_{B}} \sqrt{\frac{c_{\phi_{\rho}}^2 (K_A u_{B}+K_B u_{A})+c_{\phi_{\rho}}
		c_{\phi_{\sigma}} (2 K_A u_{B}-2 K_B u_{A})+c_{\phi_{\sigma}}^2 (K_A
		u_{B}+K_B u_{A})}{2u_{A} u_{B}}} + \mathcal{O}\left(\omega^2\right),\\
q_3&= -q_4 = \omega  \sqrt{\frac{c_{\phi_{\rho}}^2 (K_A u_{A}+K_B u_{B})+2 c_{\phi_{\rho}}
		c_{\phi_{\sigma}} (K_A u_{A}-K_B u_{B})+c_{\phi_{\sigma}}^2 (K_A
		u_{A}+K_B u_{B})}{u_{A} u_{B} \left(c_{\phi_{\rho}}^2 (K_A
		u_{B}+K_B u_{A})+c_{\phi_{\rho}} c_{\phi_{\sigma}} (2 K_A u_{B}-2
		K_B u_{A})+c_{\phi_{\sigma}}^2 (K_A u_{B}+K_B u_{A})\right)}}+ \mathcal{O}\left(\omega^3\right).
\end{align}
From Eqs.~\eqref{eq:system}, some relations between the coefficient $\epsilon_{ij}$ can be found
\begin{align}
\alpha_1=\frac{\epsilon_{cs,1}}{\epsilon_{cc,1}} &=\frac{\epsilon_{cs,2}}{\epsilon_{cc,2}}=\frac{c_{\phi_{\rho}} K_A u_{B}-c_{\phi_{\rho}} K_B u_{A}+c_{\phi_{\sigma}} K_A
	u_{B}+c_{\phi_{\sigma}} K_B u_{A}}{c_{\phi_{\rho}} K_A u_{B}+c_{\phi_{\rho}}
	K_B u_{A}+c_{\phi_{\sigma}} K_A u_{B}-c_{\phi_{\sigma}} K_B u_{A}}+ \mathcal{O}\left(\omega^1\right)\\
\alpha_3=\frac{\epsilon_{cs,3}}{\epsilon_{cc,3}} &=\frac{\epsilon_{cs,4}}{\epsilon_{cc,4}}=- \frac{c_{\phi_{\rho}}}{c_{\phi_{\sigma}}}.
\end{align}
The full solutions for the Green's function $G_{cc}$ are
\begin{equation}
\left(\begin{matrix} G_{cc}(x,\omega_n)\\G_{cs}(x,\omega_n)\end{matrix}\right)= \sum_{i=1}^{4} c_{i} e^{q_{i}x}\left(\begin{matrix} \epsilon_{cc,i}\\\epsilon_{cs,i}\end{matrix}\right)
\end{equation}
In order to find the unknown variables $c_1\epsilon_{cc,1}$ , $c_2\epsilon_{cc,2}$ , $c_3\epsilon_{cc,3}$ and $c_4\epsilon_{cc,4}$, one has to match the propagators and their first order derivatives at $x = - \frac{L}{2}$ and $ x= x' =0$. Let us observe that, by choosing $x'=0$, the Green's functions must be symmetric around $x=0$. According to this symmetry one has (let us focus on the case of $G_{cc}$)
\begin{align}
\sum_{i=1}^{4}c_{i}\epsilon_{cc,i}^{+}e^{q_i x} = \sum_{i=1}^{4}c_{i}\epsilon_{cc,i}^{-}e^{-q_i x}.
\end{align}
The coefficients $\epsilon_{cc,i}^{\pm }$ appear in the Green's function for $x>x'=0$ and $x<x'=0$, respectively. Since one has that $q_2=-q_1$ and $q_4=-q_3$, one finds the useful relations
\begin{equation}
c_1\epsilon_{cc,1}^{+} = c_2\epsilon_{cc,2}^{-}, ~~ c_2\epsilon_{cc,2}^{+} = c_1\epsilon_{cc,1}^{-}, ~~ c_4\epsilon_{cc,4}^{+} = c_3\epsilon_{cc,3}^{-},~~c_3\epsilon_{cc,3}^{+} = c_4\epsilon_{cc,4}^{-}.\label{eq:plusminus}
\end{equation}
Similar relations can be derived for the other Green's function.

The solutions in the leads ($\tilde{g}_{s_As_B}(x) = 0$) have the simple expressions
\begin{align}
G_{cc}(x,\omega_n) &= a^{(1)}_{cc} e^{-\frac{\omega }{u^L_A}\left|x\right|} + a^{(2)}_{cc} e^{-\frac{\omega}{u^L_B}\left|x\right|},\\
G_{cs}(x,\omega_n) &= a^{(1)}_{cc} e^{-\frac{\omega}{u^L_A}\left|x\right|} - a^{(2)}_{cc} e^{-\frac{\omega}{u^L_B}\left|x\right|}.
\end{align} 
The matching conditions result in a system of equations given by
\begin{align}
&G_{cc}(x = -\frac{L}{2}-\delta,\omega_n) = G_{cc}(x = -\frac{L}{2}+\delta,\omega_n),\\
&G_{cs}(x = -\frac{L}{2}-\delta,\omega_n) = G_{cs}(x = -\frac{L}{2}+\delta,\omega_n),\\
&\left\{\frac{1}{2}\left[\frac{u_{A}(x)}{K_{A}(x)}+\frac{u_{B}(x)}{K_{B}(x)}\right]\partial_x G_{cc}(x,\omega_n)+\frac{1}{2}\left[\frac{u_{A}(x)}{K_{A}(x)}-\frac{u_{B}(x)}{K_{B}(x)}\right]\partial_x G_{cs}(x,\omega_n)\right\}_{x = -\frac{L}{2}-\delta} \nonumber\\&= \left\{\frac{1}{2}\left[\frac{u_{A}(x)}{K_{A}(x)}+\frac{u_{B}(x)}{K_{B}(x)}\right]\partial_x G_{cc}(x,\omega_n)+\frac{1}{2}\left[\frac{u_{A}(x)}{K_{A}(x)}-\frac{u_{B}(x)}{K_{B}(x)}\right]\partial_x G_{cs}(x,\omega_n)\right\}_{x = -\frac{L}{2}+\delta},\\
&\left\{\frac{1}{2}\left[\frac{u_{A}(x)}{K_{A}(x)}+\frac{u_{B}(x)}{K_{B}(x)}\right]\partial_x G_{cs}(x,\omega_n)+\frac{1}{2}\left[\frac{u_{A}(x)}{K_{A}(x)}-\frac{u_{B}(x)}{K_{B}(x)}\right]\partial_x G_{cc}(x,\omega_n)\right\}_{x = -\frac{L}{2}-\delta} \nonumber\\&= \left\{\frac{1}{2}\left[\frac{u_{A}(x)}{K_{A}(x)}+\frac{u_{B}(x)}{K_{B}(x)}\right]\partial_x G_{cs}(x,\omega_n)+\frac{1}{2}\left[\frac{u_{A}(x)}{K_{A}(x)}-\frac{u_{B}(x)}{K_{B}(x)}\right]\partial_x G_{cc}(x,\omega_n)\right\}_{x = -\frac{L}{2}+\delta},\\
&\left\{\frac{1}{2}\left[\frac{u_{A}(x)}{K_{A}(x)}+\frac{u_{B}(x)}{K_{B}(x)}\right]\partial_x G_{cc}(x,\omega_n)+\frac{1}{2}\left[\frac{u_{A}(x)}{K_{A}(x)}-\frac{u_{B}(x)}{K_{B}(x)}\right]\partial_x G_{cs}(x,\omega_n)\right\}_{x = 0^-} \nonumber\\&= \left\{\frac{1}{2}\left[\frac{u_{A}(x)}{K_{A}(x)}+\frac{u_{B}(x)}{K_{B}(x)}\right]\partial_x G_{cc}(x,\omega_n)+\frac{1}{2}\left[\frac{u_{A}(x)}{K_{A}(x)}-\frac{u_{B}(x)}{K_{B}(x)}\right]\partial_x G_{cs}(x,\omega_n)\right\}_{x = 0^+} - \pi,\\
&\left\{\frac{1}{2}\left[\frac{u_{A}(x)}{K_{A}(x)}+\frac{u_{B}(x)}{K_{B}(x)}\right]\partial_x G_{cs}(x,\omega_n)+\frac{1}{2}\left[\frac{u_{A}(x)}{K_{A}(x)}-\frac{u_{B}(x)}{K_{B}(x)}\right]\partial_x G_{cc}(x,\omega_n)\right\}_{x = 0^-} \nonumber\\&= \left\{\frac{1}{2}\left[\frac{u_{A}(x)}{K_{A}(x)}+\frac{u_{B}(x)}{K_{B}(x)}\right]\partial_x G_{cs}(x,\omega_n)+\frac{1}{2}\left[\frac{u_{A}(x)}{K_{A}(x)}-\frac{u_{B}(x)}{K_{B}(x)}\right]\partial_x G_{cc}(x,\omega_n)\right\}_{x = 0^+}.
\end{align}
Then, by solving the system, one finds $c_1\epsilon_{cc,1}$ , $c_2\epsilon_{cc,2}$ , $c_3\epsilon_{cc,3}$, $c_4\epsilon_{cc,4}$ and, therefore, the propagator $G_{cc}$. The resulting conductance is given by
\begin{align}
G_S &= \frac{(g \mu_B)^2}{h} \frac{4c_{\phi_{\sigma}}^2 K_A^L K_B^L}{\left(K_A^L+ K_B^L\right)\left(c_{\phi_{\sigma}}^2+c_{\phi_{\rho}}^2\right)+2\left(K_A^L- K_B^L\right)c_{\phi_{\sigma}}c_{\phi_{\rho}}}\nonumber\\
&=\frac{(g \mu_B)^2}{h}\frac{K_A^L K_B^L (s_A-s_B)^2}{
	K_A^L s_B^2+K_B^L s_A^2}
\end{align}
where we restored Planck's constant. 
Below, we discuss some experimental setups which give rise to different values of parameters in the leads.

\subsection{Identical chains}
In the case of identical chains in the lead region ($K_A^L = K_B^L\equiv K^L$), one finds
\begin{align}
\frac{G_S}{G_S^{0}}=\frac{ (s_A-s_B)^2}{
	s_B^2+s_A^2},\label{SM:eq_idchains}
\end{align}
where $G_S^{0} \equiv K^L(g \mu_B)^2/h$ is the conductance of a single spin chain in the two-terminal configuration.

\subsection{Leads as XY chains}
When the exchange interactions along the $z$ axis are zero in the leads (XY chains), the LL parameters become $K_{A/B}^L=1$, independent of the magnetic field. In this case, the spin conductance is universal and reads
\begin{align}
G_S&= \frac{(g \mu_B)^2}{h}\frac{2(s_A-s_B)^2}{(s_A-s_B)^2+(s_A+s_B)^2}= \frac{(g \mu_B)^2}{h}\frac{(s_A-s_B)^2}{s_A^2+s_B^2}.
\end{align}

\begin{figure}[t]
	\centering
	\includegraphics[width=0.7\linewidth]{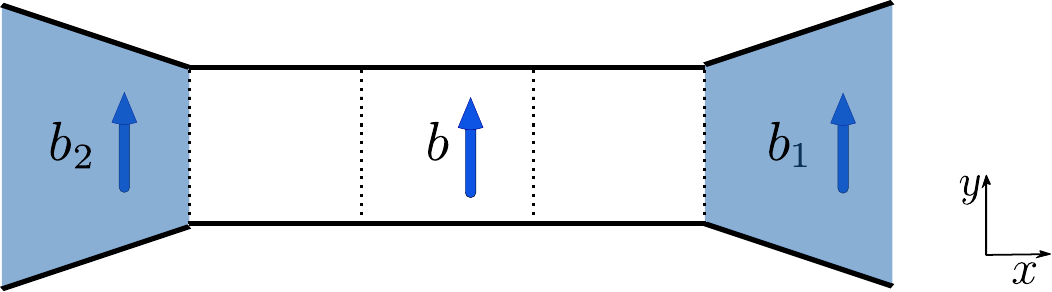}
	\caption{Setup to measure the fractional spin conductance. In the left and right leads (blue regions) different uniform magnetic fields with Zeeman energies $b^1 = b^L + \Delta b/2$ and $b^2 = b^L - \Delta b/2$ are applied. The leads are uncoupled isotropic spin-$1/2$ chains. If one assumes $b^L \sim 0 $, then the LL parameters in the chains are $K^L_A = K^L_B \equiv K^L =\frac{1}{2}$, regardless of the strength of long-range interactions.} 
	\label{fig:conductance_isotropic}
\end{figure}

\subsection{Uncoupled isotropic chains as leads}
If the distance between chains is enough to suppress the coupling between them, one can assume that the leads are isotropic chains, i.e. $J_{r}^{xy} = J_{r}^{z}$ for each $r$ (see Fig.~\ref{fig:conductance_isotropic}). In this case and for magnetic fields close to zero, i.e. $b^L = 0$, the LL  parameters are $K^L_A = K^L_B \equiv K^L =\frac{1}{2}$. As long as the chains can be described as LLs, this result is completely independent of the range of interactions. By using Eq.~\eqref{SM:eq_idchains}, one finds a universal result
\begin{align}
G_S=\frac{(g \mu_B)^2}{2h}\frac{ (s_A-s_B)^2}{
	s_B^2+s_A^2}.
\end{align}
For instance, for the process $(s_A = 2, s_B = 1)$, giving rise to fractional spin $1/3$, the spin conductance is $G_S=(g \mu_B)^2/10h $.

\section{Dynamical  structure factor}
In this section, we compute the dynamical spin structure factor in presence of the perturbation characterized by ($s_A,s_B$). The dynamical spin structure factor is defined as
\begin{equation}
\mathcal{S}(q,\omega) = \frac{1}{4\pi^2}\int dx dt~ e^{i q x} e^{-i \omega t} \left\langle \partial_x \phi_{\rho}(x,t)\partial_x\phi_{\rho}(0,0)\right\rangle
\end{equation}
since $S_z = -(1/2\pi)\partial_x \phi_{\rho}$. By expanding the cosine term up to quadratic order, the total Hamiltonian $H = H_0 + H_{\rm inter}$ expressed in this new basis is
\begin{align}
H &= \frac{1}{2\pi} \int dx~ \left\{\left[\partial_x\left(\phi_{\rho}(x),\phi_{\sigma}(x)\right)\right] N  \left[\partial_x\left(\phi_{\rho}(x),\phi_{\sigma}(x)\right)^T\right] + \left[\partial_x\left(\theta_{\rho}(x),\theta_{\sigma}(x)\right)\right] M \left[\partial_x\left(\theta_{\rho}(x),\theta_{\sigma}(x)\right)^T\right]\right\}\nonumber\\&+ \int dx~\left(\phi_{\rho}(x),\phi_{\sigma}(x)\right) \Delta \left(\phi_{\rho}(x),\phi_{\sigma}(x)\right)^T,
\end{align}
where $N$, $M$, and $\Delta$ are $2\times2$ matrices given by
\begin{align}
N &=\left(
\begin{array}{cc}
\frac{u_A}{2K_A}+\frac{u_B}{2K_B}
& \frac{u_A}{2K_A}-\frac{u_B}{2K_B}
\\
\frac{u_A}{2K_A}-\frac{u_B}{2K_B} &
\frac{u_A}{2K_A}+\frac{u_B}{2K_B} \\
\end{array}
\right),\\
M &=  \left(
\begin{array}{cc}
\frac{K_A u_A+K_B u_B}{2} & \frac{ K_A
	u_A-K_B u_B}{2}\\
\frac{ K_A
	u_A-K_B u_B}{2}& \frac{K_A u_A+K_B u_B}{2}\\
\end{array}
\right),\\
\Delta &= \tilde{g}_{s_{A}s_{B}}  \left(
\begin{array}{cc}
(s_A+s_B)^2 &s_A^2-s_B^2\\
s_A^2-s_B^2& (s_A-s_B)^2\\
\end{array}
\right).
\end{align}
Let us focus on the correlation function for the field $\phi_{\rho}$
\begin{equation}
\left \langle\phi_{\rho}(x,t)\phi_{\rho}(0,0)\right\rangle,
\end{equation} 
which can be expressed as
\begin{equation}
\left \langle\phi_{\rho}(x,t)\phi_{\rho}(0,0)\right\rangle = \frac{1}{4\pi^2}\int dq_1 d\omega_1~\int dq_2 d\omega_2~\left \langle \phi_{\rho}(q_1,\omega_1)\phi_{\rho}(q_2,\omega_2)\right\rangle e^{i q_1x - \omega_1 t}e^{i q_2x - \omega_2 t}.
\label{eq:scaling_corrq1q2}
\end{equation}
By using the above expression, one can see that the dynamical spin structure factor is
\begin{equation}
\mathcal{S}(q,\omega) = \frac{1}{4\pi^2}\left \langle \phi_{\rho}(q,\omega)\phi_{\rho}(-q,-\omega)\right\rangle .
\end{equation}
Then, by using the following relation,
\begin{equation}
\prod_{k}\left(\int~\frac{d u_k d u_k^*}{2\pi i}\right)e^{-\sum_{ij} u_i^*A_{ij}u_j+\sum_{i}h_i^* u_i+\sum_{i}u_i^* h_i}=\frac{e^{\sum_{ij}h_i^* (A^{-1})_{ij}h_j}}{\text{Det}A},
\end{equation}
we can integrate out the bosonic fields $\theta_{\rho}$ and $\theta_{\sigma}$, thus obtaining the following effective action
\begin{align}
S_{\rm eff}& = \frac{1}{2\pi} \int ~dq d\omega~  \left(\phi_{\rho}(q,\omega),\phi_{\sigma}(q,\omega)\right) \left(\omega^2 M^{-1} + q^2N+\Delta\right)  \left(\phi_{\rho}(-q,\omega),\phi_{\sigma}(-q,\omega)\right)^T.
\end{align}
Then, the correlation function for  $\phi_{\rho}$ is written as
\begin{equation}
\left\langle \phi_{\rho}(q_1,\omega_1)\phi_{\rho}(q_2,\omega_2) \right\rangle =\frac{\int \mathcal{D}\phi \mathcal{D}\theta~ \phi_{\rho}(q_1,\omega_1)\phi_{\rho}(q_2,\omega_2)~e^{- S_{\rm eff}\left[\phi_{\rho},\phi_{\sigma}\right]}}{\int \mathcal{D}\phi \mathcal{D}\theta ~e^{-S_{\rm eff}\left[\phi_{\rho},\phi_{\sigma}\right]}}.
\end{equation}
\begin{figure}[b]
	\centering
	\includegraphics[width=0.55\linewidth]{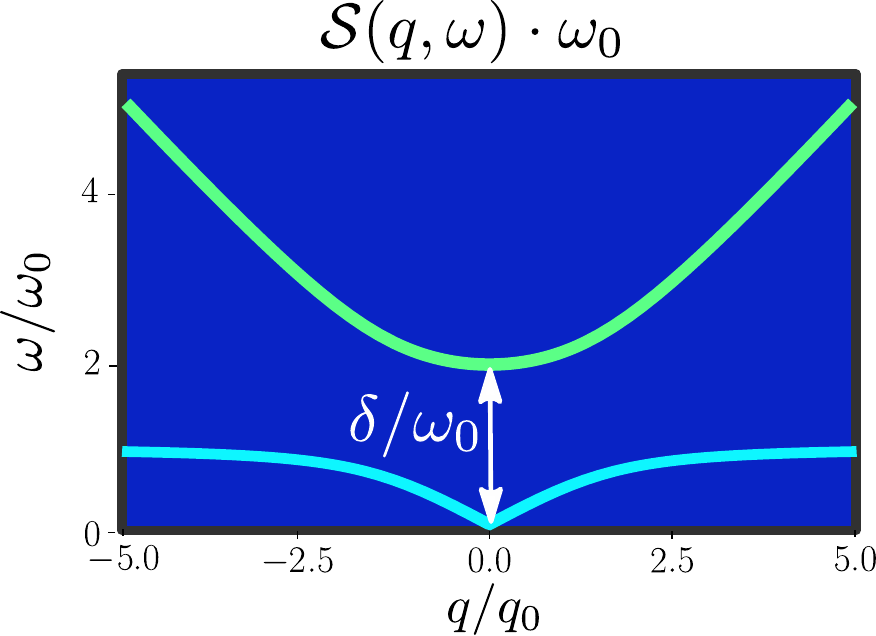}
	\caption{Color plot of the dynamical spin structure factor $\mathcal{S}(q,\omega)$ for positive energies $\omega$ and wavevector $q$ for the illustrative example of $s_A=2$, $s_B=1$ (giving fractional spin $1/3$). The function is symmetric in $q$ and $\omega$. The values of $q_0$ and $\omega_0$ are defined in the text. The dark blue background corresponds to values of $\mathcal{S}(q,\omega)$ much smaller than those of the green and light blue curves, given by the poles $\omega_{\pm}$ [Eq.~(\ref{polenegative})]. Here, $\delta=2\omega_0$ is the gap, which depends on $s_\tau$, see Eq.~\eqref{eq:SMGap}. The other parameters are chosen as $K_A=0.3$, $K_B = 0.36$, $\gamma=1.5\omega_0$ and $u_A/u_B = 3.15$.} 
	\label{fig:spinstructure}
\end{figure}One can use the following relation for gaussian integration in order to evaluate this correlation function explicitly~\cite{Giamarchi},
\begin{equation}
\left\langle u^{*}_i u_j\right \rangle = \frac{\prod_{k}\left(\int~\frac{d u_k d u_k^*}{2\pi i}\right)u^*_i u_je^{-\sum_{ij} u_i^*A_{ij}u_j}}{\prod_{k}\left(\int~\frac{d u_k d u_k^*}{2\pi i}\right)e^{-\sum_{ij} u_i^*A_{ij}u_j}} = A^{-1}_{ij}.
\end{equation}
Thus, one finds
\begin{equation}
\left\langle \phi_{\rho}(q_1,\omega_1)\phi_{\rho}(q_2,\omega_2) \right\rangle =\delta_{q_1,-q_2}\delta_{\omega_1,-\omega_2} \left[\left(\omega_1^2 M^{-1} + q_1^2N+\Delta\right)^{-1} \right]_{\rho \rho},
\end{equation}
where we selected only the diagonal component related to $\phi_{\rho}$ in the matrix. From now on, we identify for notational convenience $q_1\equiv q$ and $\omega_1 \equiv \omega$. Then, the dynamical spin structure factor $\mathcal{S}(q,\omega)$ is given by (we used that for zero temperature Matsubara frequencies  become $\omega \rightarrow i \omega$)
\begin{align}
\frac{K_B u_B \left[8\pi^2  \tilde{g}_{s_{A}s_{B}}   K_A u_A
	(s_A-s_B)^2+q^2 u_A^2-\omega ^2\right]+K_A u_A \left(q^2
	u_B^2-\omega ^2\right)}{2 \left\{q^2 \left[2  \tilde{g}_{s_{A}s_{B}}   K_A s_A^2
	u_A u_B^2+2  \tilde{g}_{s_{A}s_{B}}   K_B s_B^2 u_A^2 u_B-\omega ^2
	\left(u_A^2+u_B^2\right)\right]+\omega ^2 \left(-2  \tilde{g}_{s_{A}s_{B}}   K_A
	s_A^2 u_A-2  \tilde{g}_{s_{A}s_{B}}   K_B s_B^2 u_B+\omega ^2\right)+q^4
	u_A^2 u_B^2\right\}}. \label{polepositive}
\end{align}
The poles of this function define the spectrum of the four modes. Below, we focus on two branches defined for $\omega \ge 0$:
\begin{align}
\left(\frac{\omega_{\pm}}{\omega_0}\right)^2 = \frac{1}{2} \left[\frac{\delta ^2}{\omega_0^2}+\frac{q^2 \left(u_-^2+u_+^2\right)}{2 q_0^2
	u_- u_+}\pm \sqrt{\frac{\delta ^4}{\omega_0^4}+\frac{q^4}{q_0^4}+\frac{2 \gamma ^2
		q^2}{\omega_0^2q_0^2}}\right], \label{polenegative}
\end{align}
where we defined $u_{\pm}=\left(u_A\pm u_B\right)/2$, $\omega_0=\sqrt{\tilde{g}_{s_{A}s_{B}}u_+}$, $q_0 = \omega_0/\sqrt{4u_+u_-}$ and
\begin{align}
\gamma&=\omega_0 \sqrt{4\frac{K_A s_A^2 u_B+K_B s_B^2
		u_A}{u_A+u_B}}\label{eq:SMGamma},\\
\delta &=\omega_0 \sqrt{4\frac{K_A s_A^2 u_A+K_B s_B^2
		u_B}{u_A+u_B}}.\label{eq:SMGap}
\end{align} 
In Fig.~\ref{fig:spinstructure}, we plotted the dynamical spin structure factor for positive energies; the plot is mirror symmetric around $\omega=0$. The dispersion of coherent modes are plotted in light blue ($\omega_{-}$) and in green ($\omega_+$). The background incoherent modes are depicted in dark blue and their spectral weight is negligible compared to coherent modes $\omega_{\pm}(q)$. At $q=0$, only two modes are gapped, one for $\omega>0$ and one for $\omega<0$.
While the gap does contain the quantum numbers $s_\tau$, it seems rather difficult to extract them from measurements of the gap energy alone, since they are masked by the presence of the non-universal coupling constant $\tilde{g}_{s_{A}s_{B}}$. This is the reason
why we focus on the conductance as a way to get experimental access to the  excitations with fractional spin.

\end{document}